\documentclass[twocolumn,prd,nofootinbib,showpacs]{revtex4}
\usepackage{graphicx}
\usepackage{color}
\usepackage{amsmath,amssymb,amsfonts,dsfont}
\usepackage{multirow}
\usepackage{sistyle}
\usepackage{ulem}

\pdfoutput=1 

\newcommand{\sigmav}{$\langle\sigma v\rangle$ }

\def\lsim{\mathrel{\rlap{\lower4pt\hbox{\hskip1pt$\sim$}}
    \raise1pt\hbox{$<$}}}                
\def\gsim{\mathrel{\rlap{\lower4pt\hbox{\hskip1pt$\sim$}}
    \raise1pt\hbox{$>$}}}                

\def \aap  {A\&A}

\def \apjs  {ApJS}
\def \apjl  {ApJL}

\begin{document}

\title{\boldmath Composition of the {\it Fermi}-LAT isotropic gamma-ray background intensity: Emission from extragalactic point sources and dark matter annihilations}

\author{Mattia Di Mauro,}\email{mattia.dimauro@to.infn.it}
\affiliation{Dipartimento di Fisica, Universit\`a di Torino, via P. Giuria 1, 10125 Torino, Italy}
\affiliation{Istituto Nazionale di Fisica Nucleare, Sezione di Torino, Via P. Giuria 1, 10125 Torino, Italy}
\affiliation{LAPTh, Universit\'e de Savoie, CNRS, 9 Chemin de Bellevue, B.P.\ 110, F-74941 Annecy-le-Vieux, France}
\author{Fiorenza Donato}\email{donato@to.infn.it}
\affiliation{Dipartimento di Fisica, Universit\`a di Torino, via P. Giuria 1, 10125 Torino, Italy}
\affiliation{Istituto Nazionale di Fisica Nucleare, Sezione di Torino, Via P. Giuria 1, 10125 Torino, Italy}

\begin{abstract}
A new estimation of the isotropic diffuse $\gamma$-ray background (IGRB) observed by
the Large Area Telescope (LAT) on board the  {\it Fermi Gamma-ray Space Telescope (Fermi)} has been presented 
for 50 months of data, in the energy range 100 MeV$-$820 GeV and for different modelings of the Galactic foreground. 
We attempt here the interpretation of the  {\it Fermi}-LAT IGRB data in terms of the $\gamma$-ray unresolved emission 
from different extragalactic populations. We find very good fits to the experimental IGRB, obtained with theoretical predictions for the emission 
from active galactic nuclei and star-forming galaxies. 
In addition, we probe a possible emission coming from the annihilation of 
weakly interacting dark matter (DM) particles 
 in the halo of our Galaxy. We set stringent limits on its annihilation cross section into $\gamma$ rays,  which are about the thermal relic value 
for a wide range of DM masses. We also identify regions in the DM mass and annihilation cross section parameter space which 
can significantly improve the fit to the IGRB data. 
Our analysis is conducted within the different IGRB data sets obtained from different models for the Galactic emission, 
which is shown to add a significant ambiguity on the IGRB interpretation. 
\end{abstract}

\maketitle

\section{Introduction}
\label{sec:introcomp}
The origin and the composition of the observed isotropic diffuse $\gamma$-ray background (IGRB) 
is one of the most intriguing open problems in astrophysics. 
The existence of an isotropic diffuse component was first pointed out by OSO-3 \cite{1972ApJ...177..341K},
and then confirmed by SAS-2 \cite{1975ApJ...198..163F} and EGRET \cite{1998ApJ...494..523S} satellite-based detectors. 
The Large Area Telescope (LAT) on board the  {\it Fermi Gamma-ray Space Telescope (Fermi)} 
has first provided a measurement of the IGRB in the 200 MeV$-$100 GeV energy range \cite{2010PhRvL.104j1101A}, based on 10 months of data taking. 
This residual emission is found in the $\gamma$ ray at latitudes $ |b|>20^\circ$ sky after subtracting resolved sources, the diffuse Galactic emission,
the CR background 
and by the Sun contribution to the total all sky $\gamma$-ray emission. 
This measurement of the IGRB is consistent with a power-law spectrum with a slope of $2.41 \pm 0.05$.
Very recently, a new estimation of the IGRB has been presented by the {\it Fermi}-LAT Collaboration, 
based on 50 months of data in the range 100 MeV$-$820 GeV \cite{igrb_2014}. 
In addition, the Collaboration has provided the spectrum for the total (putative) 
extragalactic $\gamma$-ray background (EGB), given by the sum of the IGRB and the flux from resolved sources. 
For the first time, the data analysis has been conducted using three different modelings of the diffuse  
Galactic emission, which acts as a foreground in the reduction data process. 
We adopt the same definitions as in \cite{igrb_2014} for the three different cases, namely model 
A, B and C, and refer to this paper for any detail on how the Galactic foreground has been shaped. 
This new measurement  shows  a significative softening of the IGRB and EGB spectra, compatible with an energy exponential 
cutoff feature. A fit performed using a power law (with 
slope $2.32 \pm 0.02$) with an exponential cutoff (with break energy of $279 \pm 52$ GeV) has been found to properly reproduce the IGRB data \cite{igrb_2014}.

The intensity of the IGRB is usually attributed to the $\gamma$-ray emission from unresolved extragalactic sources. 
The most numerous identified source population is the blazar one,
usually divided into BL Lacertae (BL Lac) objects and flat-spectrum radio quasars (FSRQs)
according to the absence or presence of strong broad emission lines in their optical or UV spectrum, respectively. Blazars have been shown, indeed, to 
produce a significant fraction of the IGRB \cite{1998ApJ...496..752C,2000MNRAS.312..177M,narumoto2006,2007ApJ...659..958D,
Kneiske:2007jq,2011PhRvD..84j3007A,2009ApJ...702..523I,1996ApJ...464..600S,2011ApJ...736...40S,Neronov:2011kg,Ajello:2013lka,Ajellog2014}. 
Recently, it has been derived that BL Lacs contribute to 10\% of the IGRB at 100 MeV and up to 100\% at 100 GeV, 
naturally explaining the IGRB softening with increasing energy \cite{DiMauro:2013zfa}. 
A non-negligible contribution to the IGRB can also come from those active galactic nuclei (AGN) whose jet is misaligned (MAGN) along the line of sight (l.o.s.). 
The MAGN resolved by {\it Fermi}-LAT are far less numerous than blazars  due to 
the Doppler attenuation. On the other hand, because of
the
 geometrical distribution of jet emission angles, the unresolved counterpart is expected to be very numerous 
and to give thus a sizable contribution to the IGRB \cite{DiMauro:2013xta,2011ApJ...733...66I}.
A further source class for diffuse $\gamma$ rays has been identified with star-forming (SF) galaxies, 
whose unresolved flux can contribute significantly to the IGRB \cite{2012ApJ...755..164A,2010ApJ...722L.199F,Cholis:2013ena}. 
As for possible Galactic sources, pulsars have been considered as promising contributors to the IGRB. 
A very recent estimation of the $\gamma$-ray emission from high-latitude millisecond pulsars has been performed in \cite{Calore:2014oga}, based on the 
second {\it Fermi}-LAT pulsar catalog \cite{2013ApJS..208...17A} listing 117 $\gamma$-ray pulsars.  
It is found that at latitudes higher than $10^{\circ}$ at most  $1\%$ of the IGRB could be explained by this Galactic population, in agreement with
\cite{Hooper:2013nhl,2013A&A...554A..62G,Cholis:2013ena}.
In \cite{DiMauro:2013zfa}, it has been already shown how the sum of the aforementioned components can 
nicely fit the IGRB as measured by {\it Fermi}-LAT up to about 400 GeV.
Moreover, in \cite{DiMauro:2014wha} it is demonstrated that MAGN and blazar populations  can easily explain both the intensity and anisotropy of the 
IGRB \cite{2010PhRvL.104j1101A,2012PhRvD..85h3007A}.
\\
A  possible diffuse emission may also come from the annihilation of dark matter (DM) particles in the halo of the Milky Way and 
in external galaxies \cite{1990NuPhB.346..129B,Bergstrom:2001jj,Ullio:2002pj,Bottino:2004qi}. 
Indeed, an interesting DM investigation technique 
relies in the search for its stable annihilation products in the halo of galaxies, and in particular 
in the Milky Way. Assuming that the  DM in the Universe consists of weakly interacting 
massive particles (WIMPs),  one of the most promising indirect detection 
means  is the search of its annihilation into $\gamma$ rays (see Reference~\cite{Bringmann:2012ez} for a recent review). 
Upper limits on the DM annihilation cross section from the high latitudes $\gamma$-ray emission were derived, {\it e.g.}, in  
\cite{2010NuPhB.840..284C,2012PhRvD..85b3004C, Abazajian:2010zb,Calore:2013yia,Ackermann:2012rg,Ajellog2014,Ackermanna2014,Massarii2014}.

In this paper we explore i) the features that the different astrophysical components must have in order 
to fit the measured EGB and IGRB data and ii) to which extent the DM contribution can accommodate 
the {\it Fermi}-LAT data together with all the extragalactic emissions. 
The major new point of our analysis resides in a fitting procedure where all the 
contributions are left free in their theoretical uncertainty bands both in shape and normalization to fit the data. Only $a$ $posteriori$ do we check the features and mutual weight of the different components and verify their agreement with 
theoretical predictions obtained on totally independent methods.
We also extend the analysis to both the IGRB and EGB data by considering the
different modelings of the subtracted Galactic emission \cite{igrb_2014}. For the first time, we demonstrate how relevant is the role of
 the Galactic foregrounds when fitting the high-latitude $\gamma$-ray emission, and how significantly  it can affect the DM results.

\section{Astrophysical interpretation of the {\it Fermi}-LAT IGRB data}
\label{sec:astro}
A diffuse $\gamma$-ray emission has been predicted for various populations of unresolved extragalactic sources. 
We briefly review these predictions
 and then move to a statistical analysis of both the EGB and the IGRB in terms of the most relevant among these components. 

\subsection{Diffuse $\gamma$-ray emission from astrophysical sources.}
{\it FSRQs}. 
\newline
FSRQs are blazars with strong broad emission lines in their optical or UV spectrum \cite{1980ARA&A..18..321A}.
In the second {\it Fermi} catalog (2FGL) \cite{2012yCat..21990031N,secondcatalogAGN} 360 sources are classified as FSRQs, with  a redshift distribution peaked 
around $z=1$ and extending to $z=3.10$.
Reference~\cite{2010ApJ...720..435A} estimated the contribution of FSRQs to the IGRB from their source-count distribution at flux
levels
$F_{\gamma} \geq 10^{-9}$ photons cm$^{-2}$ s$^{-1}$ ($F_{\gamma}$ is the flux integrated above a threshold energy of 100 MeV)  
and found that FSRQs can contribute to the IGRB by about 10\% in the 100 MeV-100 GeV energy range. 
Recently, \cite{2012ApJ...751..108A} examined the properties of $\gamma$-ray selected FSRQs using data of the first {\it Fermi}-LAT First Source Catalog \cite{1FGLC}.
The spectral energy distribution (SED) of the detected FSRQs shows some curvature, with a peak in the 10 MeV$-$10 GeV range, 
followed by a decrease leading to undetectable fluxes at energies higher than 30 GeV. 
The modeled SED and luminosity function lead to a predicted contribution of the FSRQs to the IGRB of about $8\%-11 \%$ below 10 GeV, which drops to negligible 
percentages for higher energies. 
Indeed, due to the redshift distribution, their SED and the absorption of $\gamma$ rays by the extragalactic background light (EBL), the FSRQs 
are expected to give a negligible contribution to the IGRB above 10 GeV.
\\
\\
{\it BL Lacs}. 
\newline
BL Lacs are blazars characterized by the absence of strong broad emission lines in their optical or UV spectrum \cite{1980ARA&A..18..321A}.
In the 2FGL catalog \cite{2012yCat..21990031N,secondcatalogAGN} 423 sources are classified as BL Lacs. 
Given the absence of broad emission lines it is quite difficult to derive the  redshift of BL Lacs, which however peaks around  $z=0.2$ and extends to $z=1.5$. 
An analysis based on the source-count distribution  \cite{2010ApJ...720..435A} has estimated the contribution of unresolved BL Lacs to the IGRB 
at about 10\% level in the 100 MeV$-$100 GeV. 
According to \citep{1995ApJ...444..567P,2010ApJ...716...30A}, blazars can be classified with respect to the position of the synchrotron-peak frequency $\nu_S$. 
Low-synchrotron-peaked blazars (LSP) are defined for observed peak frequency in the far infrared or infrared band ($\nu_S < 10^{14}$ Hz), 
intermediate-synchrotron-peaked (ISP) for $\nu_S$  bracketed in the
near IR to ultraviolet (UV) frequencies ($10^{14} \,\,\rm{Hz}\leq \nu_S <
10^{15}$ Hz), and  high-synchrotron-peaked (HSP) when the peak frequency
is found at  UV or higher energies ($\nu_S \geq 10^{15}$ Hz). 
Recently, \cite{DiMauro:2013zfa,Ajello:2013lka} derived two independent theoretical analyses of the unresolved emission from BL Lacs.
In Reference~\cite{Ajello:2013lka} a sample of 211 sources of the 1FGL catalog \cite{Abdo:2010ru} has been used 
to determine the luminosity function of BL Lacs and its evolution with redshift. 
They find that the contribution of this source population corresponds to about $7\%-10\%$ of the integrated IGRB.
In Reference~\cite{DiMauro:2013zfa} the 2FGL catalog of AGN has been used \cite{secondcatalogAGN} together with the 1FHL catalog \cite{TheFermi-LAT:2013xza} and all the available data from the Cherenkov Telescopes Array collected in the TeV catalog \cite{tevcat}.
It is found that the BL Lacs overall contribution to the IGRB is about 11\%, while the differential spectrum 
increases from about 10\% of the measured IGRB at 100 MeV to about~100\%  at 100 GeV. Above 100 GeV, the predicted flux explains the softening of the measured 
spectrum as due to the EBL attenuation.
\\
\\
{\it MAGN}. 
\newline
The $\gamma$-ray emitting MAGN sources have been studied in \cite{2011ApJ...733...66I,DiMauro:2013xta}, 
relying on a correlation between the luminosities  in the radio and $\gamma$-ray frequencies. 
Reference~\cite{2011ApJ...733...66I} finds that the unresolved counterpart of this AGN population accounts for about $25\%$ of the IGRB.
On the other hand, \cite{DiMauro:2013xta} finds that  this population can explain about the $20\%-30\%$ of the IGRB at all energies, embedded in an uncertainty band of a factor ten. This uncertainty is associated to the smallness of the resolved sample and to the errors in the $\gamma$-ray and radio 
luminosity correlation.
\\
\\
{\it SF galaxies.}  
\newline
For galaxies where star formation takes place, a guaranteed $\gamma$-ray emission arises from the decay 
of neutral pions produced in the inelastic interactions of the cosmic rays with the interstellar medium, just as in the Milky Way. Another possible
source of  $\gamma$ rays is due to electrons interactions with the gas 
(bremsstrahlung) or with the interstellar radiation fields through inverse Compton scattering (ICS). 
The {\it Fermi}-LAT has detected nine individual galaxies, four of which reside within the Local Group 
(the SMC, LMC, M31, and Milky Way) while  five  are more distant  \cite{2012ApJ...755..164A,2010ApJ...709L.152A,2011ApJ...734..107L}. 
Many more galaxies have been detected at IR wavelengths, and correlations between the two bands are speculated in order to 
predict the $\gamma$-ray diffuse emission for the unresolved SF galaxy population. 
Because of  the paucity of statistics, the SF $\gamma$-ray average 
spectrum is  difficult to firmly establish. In order to take into account expected differences in the spectra of quiescent and starburst galaxies, 
Reference~\cite{2012ApJ...755..164A} proposes two different models. 
The first refers to Milky Way-like SF galaxies (model MW), 
while the second one assumes a power-law spectrum, as  exhibited by the {\it Fermi}-LAT detected starburst galaxies (model PL). 
The two predictions differ in particular  above 5 GeV, where the MW model softens significantly. 
At 100 GeV the PL model exceeds the MW 
model by one order of magnitude by about a factor of 10.  
The estimates for the diffuse emission from SF galaxies cover from  4\% to 23\% of the IGRB intensity above 100 MeV.
Further predictions have been derived in \cite{2010ApJ...722L.199F,2013ApJ...773..104C}, 
obtaining in particular  more intense low-energy spectra. 
Very recently, Reference~\cite{2014JCAP...09..043T} has evaluated the diffuse spectrum from unresolved SF galaxies putting emphasis on 
the neutrino counterpart and finding a diffuse $\gamma$-ray emission
from normal and starburst SF galaxies comparable to the one obtained for the MW model in Reference~\cite{2012ApJ...755..164A}.  
\\
Other possible $\gamma$-ray emission mechanisms may arise from truly diffuse processes 
(i.e~extended source contributions like from ultrahigh energy cosmic rays,  nearby galaxy clusters
or gravitationally induced shock waves during structure formation, see Reference~\cite{2012PhRvD..85b3004C} and references therein).  
The uncertainties associated to these predictions are still quite large and encompass a subdominant contribution 
\cite{2011PhLB..695...13B,2004APh....20..579G,2014arXiv1410.8697Z,Calore:2013yia}. For these reasons, they are discarded from now on. 
\\
For the sake of clarity, we recall here that the resolved emission from an astrophysical source may be computed as a function of the photon energy $E$ as
\begin{eqnarray}
\label{eq:diff}
   \displaystyle \frac{dN}{dE} (E) &=& \int^{\Gamma_{{\rm max}}}_{\Gamma_{{\rm min}}} 
   d\Gamma \int^{z_{{\rm max}}}_{z_{{\rm min}}} dz \int^{L^{{\rm max}}_{\gamma}}_{L^{{\rm min}}_{\gamma}} dL_{\gamma} \,\Theta_{\gamma}(z,\Gamma,L_{\gamma}) \cdot  \nonumber \\
 &\cdot& \frac{d F_{\gamma}} {dE} (E,\Gamma) e^{-\tau_{\gamma\gamma}(E,z)} \omega(F_{\gamma}),
\end{eqnarray}
where $\Gamma$ is the spectral index of a typical power-law SED in the range 0.1$-$100 GeV, $z$ is the redshift, $L_{\gamma} $ is the $\gamma$-ray luminosity integrated in the range 0.1$-$100 GeV and 
$\Theta_{\gamma}(z,\Gamma,L_{\gamma})$ is the space density of a given population \cite{DiMauro:2013zfa}. 
The term  $d F_{\gamma}/dE$ is the intrinsic photon flux, while the exponential factor takes into account the absorption on the EBL 
(see \cite{DiMauro:2013xta,DiMauro:2013zfa,2012ApJ...751..108A} for all the details). 
Finally, and  quite importantly for the aim of the present discussion, the term $\omega(F_{\gamma})$ represents the 
${\it Fermi}$-LAT efficiency \cite{2010ApJ...720..435A} for a source with a flux $F_{\gamma}$ to be detected. 
For the computation of the diffuse flux from the {\it unresolved} counterpart of a given source population, 
the efficiency $\omega(F_{\gamma})$  in Eq.~\ref{eq:diff} must by shifted into $\omega_{unres}(F_{\gamma})\equiv(1-\omega(F_{\gamma}))$, 
while for the total (resolved plus unresolved) emission the efficiency is set to 1. 
For the BL Lac and MAGN populations we use the estimation for $\omega(F_{\gamma})$ derived in  \cite{DiMauro:2013xta},  
while for the SF galaxies and FSRQs it has been used the efficiency derived in \cite{2010ApJBlazarFermi}.  
For the absorption on the EBL we adopt the model in \cite{2010ApJ...712..238F} consistent with the recent observations of {\it Fermi}-LAT and H.E.S.S. Collaboration \cite{Ackermann:2012sza,Abramowski:2012ry}. 
\begin{figure*}
\centering
\includegraphics[width=\columnwidth]{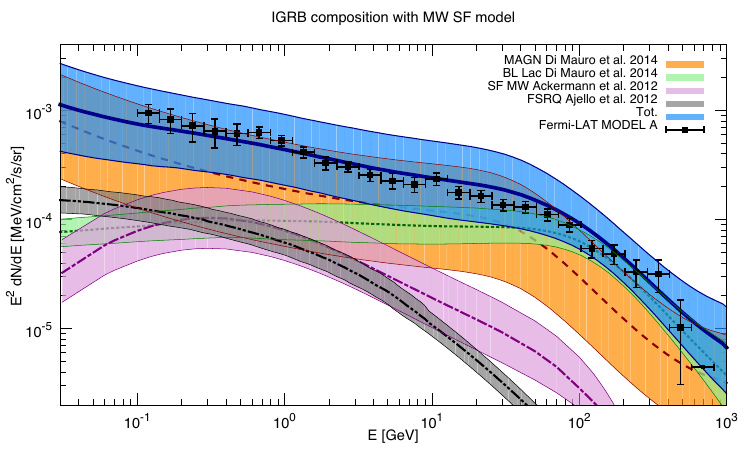}
\includegraphics[width=\columnwidth]{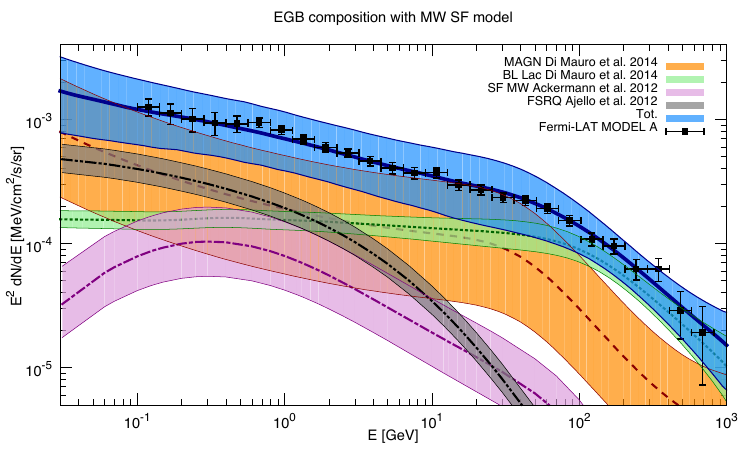}
\includegraphics[width=\columnwidth]{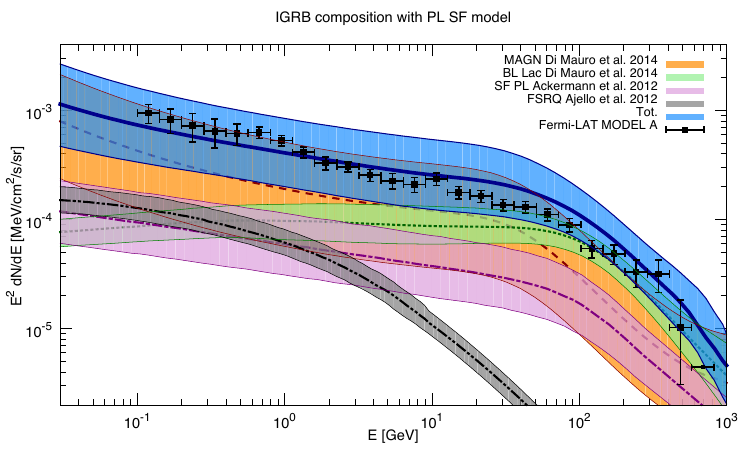}
\includegraphics[width=\columnwidth]{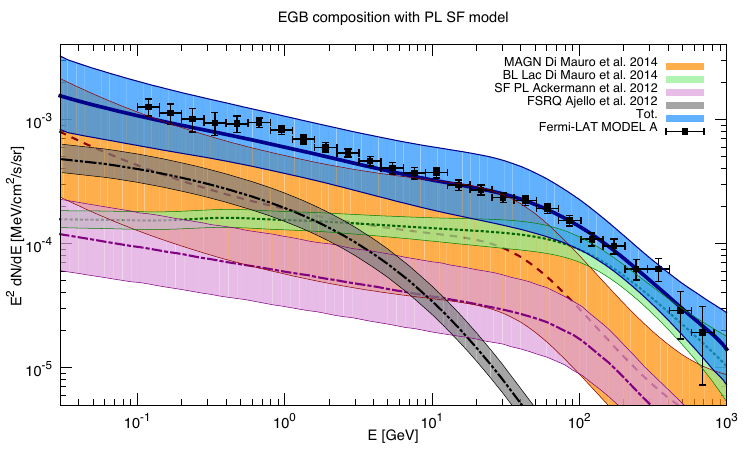}
\caption{Left (right) panels: $\gamma$-ray emission from unresolved (total=unresolved+resolved) sources, along with 
data for the IGRB (EGB) \cite{igrb_2014}. Lines and relevant uncertainty bands represent the contribution from the following source populations: 
orange dashed for MAGN,  green dotted for BL Lacs, grey double dot-dashed for FSRQs, purple dot-dashed for SF galaxies, and blue solid for the sum of all 
the contributions. Upper (lower) panels refer to the MW (PL) model for SF galaxies.  
Experimental results have been obtained for the Galactic foreground model A. }
\label{fig:igrb_egb_theo}
\medskip
\end{figure*}

In Fig.~\ref{fig:igrb_egb_theo} we report the {\it Fermi}-LAT data for the IGRB and EGB fluxes \cite{igrb_2014},  together with the emissions predicted for the populations discussed above. 
For each population, we plot the associated
uncertainty band, as evaluated in the relevant papers:
BL Lacs in  \cite{DiMauro:2013zfa}, FSRQs in \cite{2012ApJ...751..108A}, MAGN in \cite{DiMauro:2013xta} 
and  SF galaxies (both MW and PL models) in \cite{2012ApJ...755..164A}. 
The left panels of the figure show the different contributions from unresolved sources to the IGRB and their estimated uncertainty bands, along with the 
${\it Fermi}$-LAT data. The upper (lower) panel refers to the MW (PL) model for SF galaxies. The sum of each component is 
depicted by the blue line and relevant band, and shows clearly that the IGRB data are remarkably well explained by diffuse emission 
from unresolved AGN and SF galaxies, with negligible effect induced by different models for the SF galaxy emission. 
The right panels show the effect of including the resolved sources along, with the EGB data. 
A simple by-eye inspection shows that the addition of the resolved sources to the theoretical models keeps the very good
agreement with the experimental data. 

\begin{figure*}
\centering
\includegraphics[width=1.2\columnwidth]{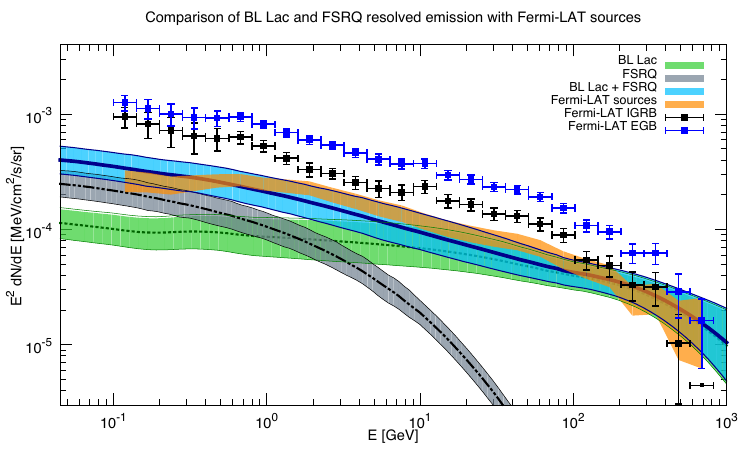}
\caption{$\gamma$-ray mission from resolved sources. The dotted green line and uncertainty band corresponds to the prediction for 
BL Lacs, the double dotted-dashed gray line and band are for FSRQs, while their sum is depicted by the solid blue line and relevant uncertainty band. 
The sources detected by the {\it Fermi}-LAT are represented by the orange band \cite{igrb_2014}.  The upper blue (lower black) data refer to the EGB (IGRB)
{\it Fermi}-LAT data, and include the emission from the resolved and unresolved (unresolved only) sources. Experimental results 
have been obtained for the 2FGL blazars within the Galactic foreground model A.}
\label{fig:sumres}
\medskip
\end{figure*}

For the sake of completeness, in Fig.~\ref{fig:sumres} we compare the emission predicted for the resolved extragalactic sources 
along with the relevant {\it Fermi}-LAT measurements. 
Since the sample of detected SF galaxies and MAGN is negligible with respect to FSRQ and BL Lac objects, 
we plot only the $\gamma$-ray flux coming from blazars. The models are derived following the 
above prescription for the required efficiency. 
The comparison between the {\it Fermi}-LAT  data on all the resolved sources (orange band in \cite{igrb_2014}) and the predictions 
(blue solid line and band) confirms that also the resolved part of the high-latitude diffuse emission 
is well explained by the phenomenological models assumed in the present work. In Fig.~\ref{fig:sumres} it is also clearly visible that the resolved sources contribute by a fraction of 20$-$30\% of the total high-latitude emission for almost all the energy range explored by the LAT. 

\subsection{Astrophysical interpretation of the IGRB data}
\label{fit:bgd}
In this section, we determine to which extent the diffuse emission coming from the 
various populations discussed in Sec.~\ref{sec:astro} can explain the IGRB data. 
As a consistency check, we will repeat the same procedure to the EGB spectrum. 
In all the following analysis we will assume the predictions for the diffuse $\gamma$-ray emission illustrated in Fig.\ref{fig:igrb_egb_theo}, namely: BL Lacs  derived in \cite{DiMauro:2013zfa}, FSRQs in \cite{2012ApJ...751..108A}, MAGN in \cite{DiMauro:2013xta} and 
 SF galaxies (both MW and PL models)  as in \cite{2012ApJ...755..164A}.
The idea is to perform a fit to the IGRB data with these contributions considered 
within their predicted theoretical uncertainties. Our aim is to prove that the extragalactic diffuse emission from known source populations explains the observed IGRB spectrum or, at variance, that an additional, more exotic component is needed to better explain the data.  
\\
We have proceeded with a $\chi^2$ fitting method with $M$ free parameters $\vec{\alpha}=\{\alpha_1, ..., \alpha_M\}$ identified on the basis of the physical properties 
of the fluxes of the various contributing populations. On a general basis, we have defined
\begin{equation}
     \label{eq:chifunc}
     \chi^2(\vec{\alpha}) =   \sum^{N}_{j=1} \frac{\left(\frac{dN}{dE}(\vec{\alpha},E_j) - \frac{dN_{\rm exp}}{dE}(E_j)\right)^2}{\sigma^2_j} + 
      \sum_{i=1}^M \frac{(\alpha_i - \bar{\alpha_i})^2}{\delta^2_i},
\end{equation}
where  $dN_{\rm exp}/dE(E_j)$ and $\sigma_j$ are the experimental flux and 1-$\sigma$ error running on N energy bins, and 
$dN/dE(\vec{\alpha},E_j)$ is the total theoretical $\gamma$-ray emission evaluated within the $\vec{\alpha}$ set of free parameters
and in each energy bin $E_j$. 
The parameters $\bar{\alpha_i}$ and $\delta_i$ correspond to the average and 1-$\sigma$ uncertainty values, respectively, found 
for the theoretical predictions of the various source populations (see relevant papers, namely \cite{DiMauro:2013zfa,2012ApJ...751..108A,DiMauro:2013xta,2012ApJ...755..164A}). 
The second term in the $\chi^2$ function of Eq.~\ref{eq:chifunc}  takes into account the uncertainties on the theoretical modeling, 
disfavoring values of $\alpha_i$ far from $\bar{\alpha_i}$, with the weight $\delta_i$. 
\\
For the choice of the free parameters of the fit, we can reason as follows.
The $\gamma$-ray emission from MAGN  strongly depends on the correlation between the radio and $\gamma$ luminosities \cite{DiMauro:2013xta}, which induces 
an uncertainty of about one order of magnitude in the estimated flux reaching Earth. The uncertainty in the $\gamma$-ray luminosity acts essentially as a scaling factor 
of the flux, as clearly visible in Fig.~7 of Reference~\cite{DiMauro:2013xta}. The global shape of the $\gamma$-ray flux is driven by the power-law index $\Gamma$ of the MAGN population SED, 
for which it was adopted a Gaussian distribution around the average value $\bar{\Gamma}$=2.37 with 1-$\sigma$ dispersion of 0.32, and integrated according to Eq.~\ref{eq:diff}. 
We have therefore chosen to translate  the uncertainty on this source population into a normalization factor
with respect to the average flux (i.e. the solid line in  Fig.~7 of Reference~\cite{DiMauro:2013xta}). 
For the blazar population, the luminosity function has been derived directly from the $\gamma$-ray catalogs. The uncertainty in the luminosity function induces an uncertainty on the 
 $\gamma$-ray flux less than a factor of 2. For the BL Lacs population, the uncertainty slightly increases due to the energy cutoff assumed for the SED of the HSP BL Lacs. 
As for the case of MAGN, the uncertainty on the  diffuse $\gamma$-ray flux for both populations has been translated into an overall renormalization with respect to the average 
contribution declared in \cite{DiMauro:2013zfa} for BL Lacs 
and in \cite{2012ApJ...751..108A} for FSRQs. In particular, for BL Lacs we treat separately the LISP (LSP and 
ISP, see \cite{DiMauro:2013zfa} for details) and HSP populations, thus introducing two normalization parameters in the fit to the IGRB. 
Finally, the SF galaxies unresolved emission depends from an IR-$\gamma$ luminosity correlation and the 
assumed SED shape (MW or PL model). Also in this case,  the variation of the  IR-$\gamma$ correlation gives an overall 
uncertainty band of about a factor 4, which can again be described by a scaling factor \cite{2012ApJ...755..164A}. 
We take effectively into account the possible uncertainties brought by different SED parameterizations by discussing separately the MW and PL models. 
\\
\begin{table}[t]
\center
\scalebox{0.9}{
\begin{tabular}{|c|c|c|c|c|c|}
\hline
 IGRB  &   BL Lac  &   FSRQ    &  MAGN     &  SF(MW)  &  $\chi^2_{\rm{IGRB}}$   \\
\hline
MODEL A  &   $0.90\pm0.05$   &  $1.03\pm0.06$   &  $0.83\pm0.07$   & $1.18\pm0.17$   & 34.4   \\
MODEL B  &   $0.96\pm0.05$   &  $1.02\pm0.06$   &  $1.49\pm0.09$   & $1.22\pm0.17$   & 26.5   \\
MODEL C  &   $0.94\pm0.05$   &  $1.01\pm0.06$   &  $0.88\pm0.07$   & $1.07\pm0.17$   & 16.4   \\
\hline
 IGRB  &   BL Lac  &   FSRQ    &  MAGN     &  SF(PL)  &  $\chi^2_{\rm{IGRB}}$   \\
\hline
MODEL A  &   $0.85\pm0.05$   &  $1.04\pm0.06$   &  $0.79\pm0.08$   & $0.94\pm0.09$   & 64.1   \\
MODEL B  &   $0.91\pm0.05$   &  $1.04\pm0.06$   &  $1.44\pm0.09$   & $0.98\pm0.09$   & 45.9   \\
MODEL C  &   $0.88\pm0.05$   &  $1.02\pm0.06$   &  $0.83\pm0.07$   & $0.95\pm0.09$   & 33.3   \\
\hline
\end{tabular}
}
\caption{Best fit on the {\it Fermi}-LAT IGRB and 1-$\sigma$ values for the BL Lacs, FSRQ, MAGN, SF (MW model for the first three rows, PL model for the last ones) 
normalization factors, reported with respect their relevant theoretical average values. 
The last column reports the $\chi^2$ value. model A, B and C refer to the {\it Fermi}-LAT data obtained within 
three different modelings of the Galactic foreground \cite{igrb_2014}.}
\label{tab:fitbackigrb}
\end{table}
\begin{table}[t]
\center
\scalebox{0.9}{
\begin{tabular}{|c|c|c|c|c|c|}
\hline
 EGB  &   BL Lac  &   FSRQ    &  MAGN     &  SF(MW)  &  $\chi^2_{\rm{EGB}}$   \\
\hline
MODEL A  &   $1.00\pm0.05$   &  $1.06\pm0.06$   &  $1.33\pm0.10$   & $1.15\pm0.17$   & 20.0   \\
MODEL B  &   $1.03\pm0.05$   &  $1.06\pm0.06$   &  $2.00\pm0.11$   & $1.18\pm0.17$   & 33.0   \\
MODEL C  &   $1.03\pm0.05$   &  $1.02\pm0.06$   &  $1.38\pm0.10$   & $1.06\pm0.17$   & 12.6   \\
\hline
 EGB  &   BL Lac  &   FSRQ    &  MAGN     &  SF(PL)  &  $\chi^2_{\rm{EGB}}$   \\
\hline
MODEL A  &   $0.97\pm0.04$   &  $1.09\pm0.06$   &  $1.26\pm0.09$   & $1.00\pm0.09$   & 29.6   \\
MODEL B  &   $0.99\pm0.04$   &  $1.09\pm0.06$   &  $1.94\pm0.11$   & $1.07\pm0.17$   & 38.4   \\
MODEL C  &   $1.03\pm0.04$   &  $1.05\pm0.06$   &  $1.31\pm0.10$   & $1.01\pm0.17$   & 16.1   \\
\hline
\end{tabular}
}
\caption{The same as in Table~\ref{tab:fitbackigrb} but for the fit to the EGB data.}
\label{tab:fitbackegb}
\end{table}
\noindent
We end up with five free parameters ($M=5$), corresponding to five effective normalizations of the theoretical contributions to the IGRB, to be included in the $\chi^2$ procedure 
described in Eq.~\ref{eq:chifunc}.  
The fit has been performed on the IGRB data and for all the three different modelings (A, B, C) of the Galactic foreground \cite{igrb_2014}. 

\begin{figure*}
\centering
\includegraphics[width=\columnwidth]{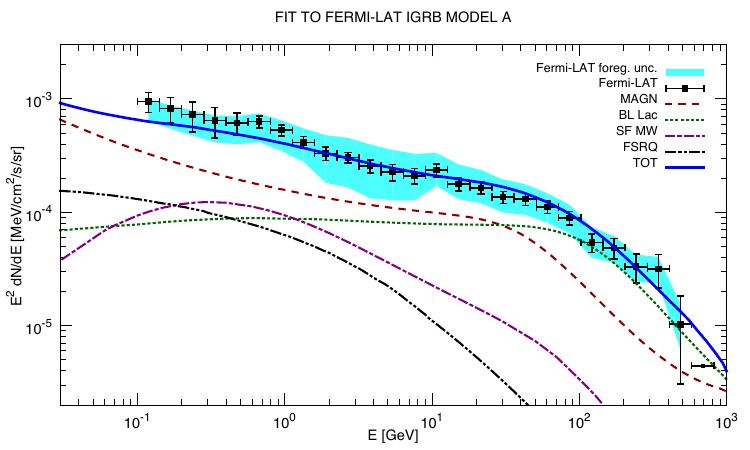}
\includegraphics[width=\columnwidth]{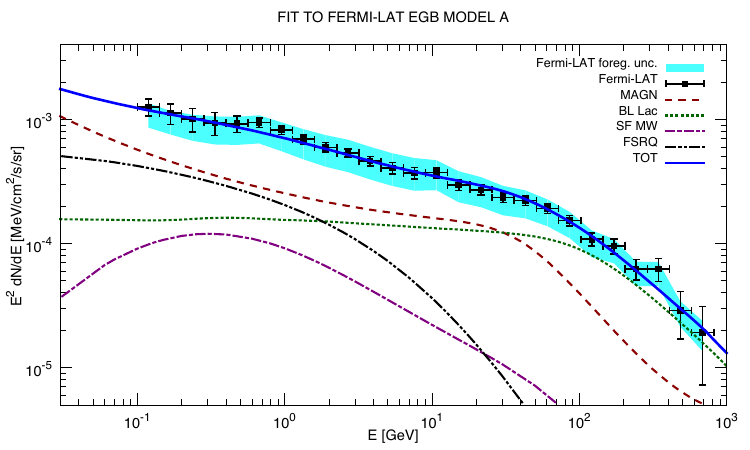}
\includegraphics[width=\columnwidth]{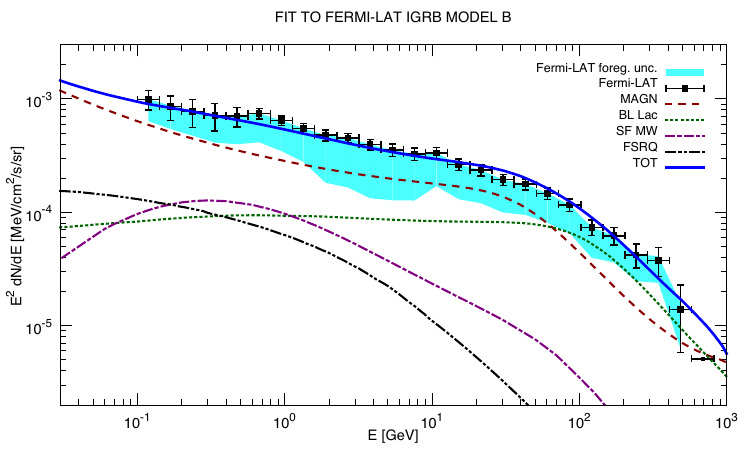}
\includegraphics[width=\columnwidth]{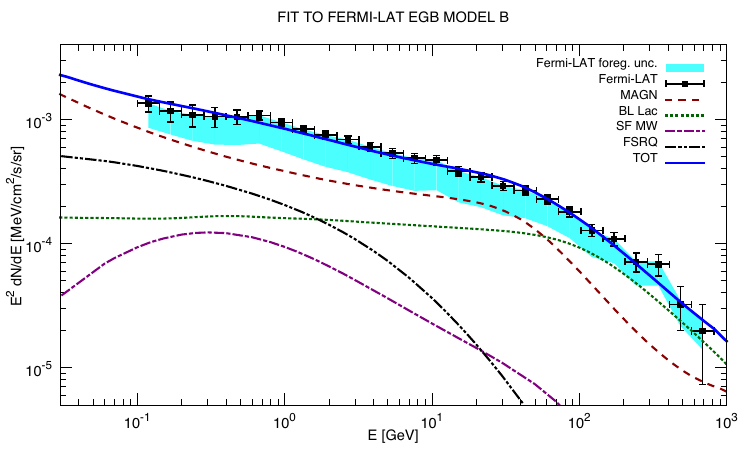}
\includegraphics[width=\columnwidth]{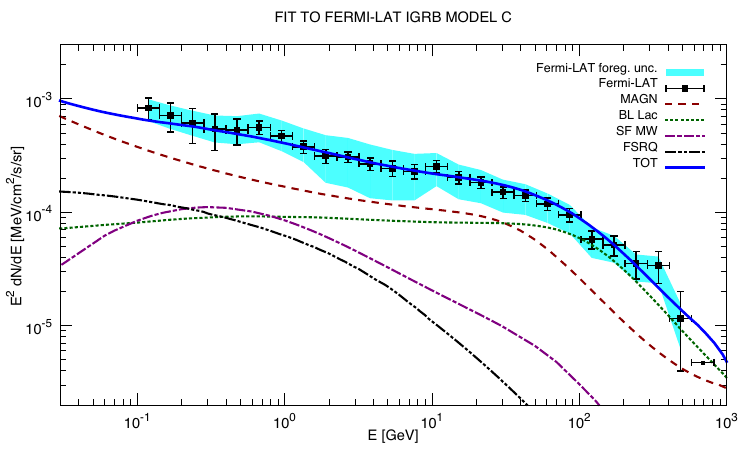}
\includegraphics[width=\columnwidth]{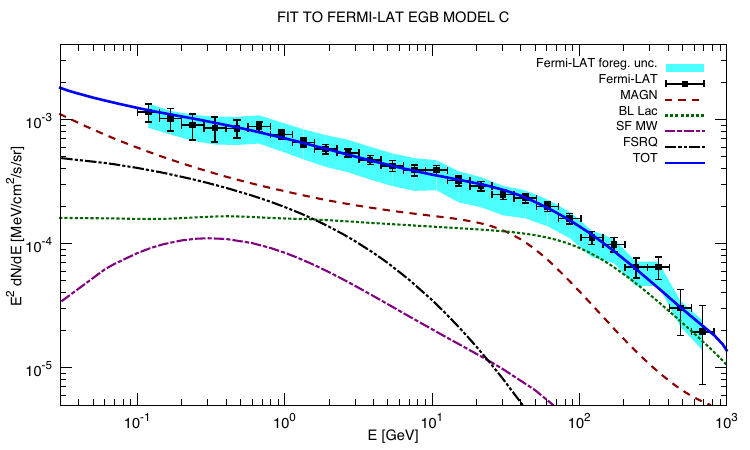}
\caption{The differential $\gamma$-ray flux obtained for the extragalactic sources (AGN+SF) according to the best fit to the IGRB and EGB data in Tabs.~\ref{tab:fitbackigrb} and \ref{tab:fitbackegb}. The BL Lac (dotted green line), MAGN 
(dashed brown), SF (dot-dashed purple line), FSRQ population (dot-dot-dashed black line), the sum of AGN and SF (solid blue line)  are shown. 
We display with a cyan band also the uncertainty associated to the {\it Fermi}-LAT foreground model \cite{igrb_2014}.}
\label{fig:backfit}
\medskip
\end{figure*}

Our main results are summarized in Table~\ref{tab:fitbackigrb}, where we report the value of the best fit and 1-$\sigma$ errors for the normalization factors of the different astrophysical 
populations. As for the BL Lacs, we report only the results for the HSP population, since the LISP (which are included in the fit) are a negligible contribution.  
The values of the normalizations are reported here with respect to their average value, fixed for each population as declared in the relevant papers.
That is to say, a value of 0.90 found for BL Lacs means that
the best fit to the IGRB is obtained by decreasing the contribution of the BL Lac population by 10\% with respect the average flux found in Reference~\cite{DiMauro:2013zfa}. The same 
reasoning holds for all the other contributions. The analysis has been conducted separately for the two models (MW and PL) of the SF galaxies' emission.
The last column reports the results on the $\chi^2$, evaluated on 25 data points and for 5 free parameters. 
The main result we can read from Table~\ref{tab:fitbackigrb} is  that it is possible to explain quite well the IGRB spectrum in terms of AGN and SF galaxy emission. 
The diffuse $\gamma$-ray fluxes (namely for each of the four astrophysical components) which give the  best fit to IGRB data are very close to  their average value
found on independent phenomenological grounds. 
The theoretical predictions for the $\gamma$-ray emissions of BL Lacs, FSRQs, MAGN and SF galaxies are found to explain very well the high-latitude IGRB data with no need for significant adjustments. 
A better agreement with the IGRB data, however, is provided by the MW modeling of the SF galaxy emission. 
The Galactic foreground emission adopted for extracting the IGRB data has a remarkable relevance on the goodness of the fit. 
The fit to model C data  always provides better results, while
the data for model A are worse accommodated by the extragalactic diffuse emissions considered in our analysis. 
\\
For the sake of completeness, we have repeated the same analysis on the EGB data (26 data points, $M=5$). The results are summarized in Table~\ref{tab:fitbackegb}. 
The inclusion of resolved sources leads to a slightly better fit of the data for foreground models A and C (which is again the best fitted among the three ones). 
Beside this observation, the comments about the IGRB fit  also hold for the EGB fit. 
In order to illustrate these results, we plot in Fig.~\ref{fig:backfit} the fluxes corresponding to the best-fit parameters listed in Table~\ref{tab:fitbackegb}, both to the IGRB (left panels)
and to the EGB (right panels). The three rows correspond to the {\it Fermi}-LAT data obtained with model A, B and C for the Galactic foreground. 
Along with the data, we plot (as a light-blue band) the uncertainty due to the diffuse Galactic emission uncertainty as declared in Reference~\cite{igrb_2014}. 
The different curves in each panel represent the flux due to BL Lacs (LISP+HSP, short dashed), FSRQs (double dot-dashed),  MAGN (long dashed), SF galaxies (dot-dashed) and their 
sum (solid line), for the normalization values given in Tabs.~\ref{tab:fitbackigrb} and \ref{tab:fitbackegb}. As already demonstrated with the $\chi^2$ analysis, the IGRB and EGB data  are well interpreted by
the contribution of extragalactic sources, namely blazars, MAGN and galaxies with stars in formation. 
Given this very good agreement with the IGRB and EGB data, it is a matter of fact that the different contributions add in a way their sum can be described by a unique power law with an exponential cutoff.  The high energy slope, included the cutoff, is indeed given by the HSP BL Lacs. 
The energy cutoff in the predicted  HSP BL Lacs energy spectrum is due to the  attenuation by the EBL \cite{DiMauro:2013zfa}. 
It is also visible in Fig.~\ref{fig:backfit} that the 
MW model for SF galaxies gives a better fit to the IGRB and  EGB {\it Fermi}-LAT data.
Indeed, the model PL for SF galaxies is in some tension with the model A and B IGRB data (see Tabs.~\ref{tab:fitbackigrb} and \ref{tab:fitbackgammafree}), and in a milder 
measure also with the EGB data (see Tabs.~\ref{tab:fitbackegb} and \ref{tab:fitbackgammafree}).  
\\
\begin{table}[t]
\center
\scalebox{0.9}{
\begin{tabular}{|c|c|c|c|c|}
\hline
 $\tilde{\chi}^2$  &  IGRB (MW)  &   EGB (MW)    &  IGRB (PL)    &  EGB (PL)   \\
 \hline
MODEL A  &   1.72; 1.56   &   0.95; 1.02     &  3.20; 2.54    &   1.41; 1.36    \\
MODEL B  &   1.33; 1.32   &   1.57; 1.72     &  2.30; 1.96    &   1.83; 2.06    \\
MODEL C  &   0.82; 0.84   &   0.60; 0.60     &  1.67; 0.95    &   0.77; 0.84    \\ 
\hline
\end{tabular}
}
\caption{Fits to the IGRB and the EGB {\it Fermi}-LAT data adding the SED power-law index $\Gamma$ for each AGN population as a free parameter.  
We report the reduced chi-square value $\tilde{\chi}^2 = \chi^2/{\rm d.o.f.}$ for the fits performed  using only the normalizations as free parameters ($M=5$, left 
numbers in each column) or varying also the  slope of the spectra ($M=9$, right numbers in each column). 
The values have been derived for the three Galactic foreground models considering the EGB and IGRB {\it Fermi}-LAT data, and both the MW and PL SF galaxy models.}
\label{tab:fitbackgammafree}
\end{table}
As a further analysis of the IGRB (and EGB) data, we have explored the possibility that the theoretical predictions adopted up to now may indeed 
be affected by an additional energy shape uncertainty. This option has been implemented by varying also the spectral  power index $\Gamma$ of the SEDs of each AGN source population: 
LISP and HSP BL Lacs, FSRQs and MAGN. As already noticed, the uncertainties inherent the SF galaxies SED modeling have been treated by performing all our previous analysis
in the MW and PL models.
For each population the free spectral index $\Gamma$ is assumed to be distributed according to a Gaussian with the same dispersion as in the 
previous analysis, while the central value is let free to move in the 1-$\sigma$ band (the same adopted in the relevant literature). The parameter $M$ indicating the number of free parameters in Eq.~\ref{eq:chifunc} is now $M=9$ (five normalizations and four spectral index). 
The results are summarized in Table~\ref{tab:fitbackgammafree}, where we report the reduced $\tilde{\chi}^2 = \chi^2/{\rm d.o.f.}$ (d.o.f. is the number of degrees of freedom) on both the IGRB and the EGB, for all the three Galactic foreground cases, and for both MW and PL SF galaxy emission. 
\newline
An independent indication of the extragalactic nature of the IGRB has been obtained in \cite{Fornengo:2014cya}. 
They have reported the measurement of the angular power spectrum of cross-correlation between the unresolved component of the Fermi-LAT $\gamma$-ray 
maps and the CMB Planck lensing potential map. This result excludes a dominant contribution to the IGRB by Galactic source populations.

\section{Adding DM annihilation to the  interpretation of the {\it Fermi}-LAT IGRB data}
\label{sect:dm}
In addition to the emission from extragalactic source populations discussed in the previous section, 
a possible contribution from DM pair annihilation, both in the halo of the MW and in external halos,
can be hidden in the photons of the IGRB
\cite{Bergstrom:2001jj,Ullio:2002pj,Taylor:2002zd,Papucci:2009gd,Cirelli:2009dv,Baxter:2010fr,Abazajian:2010zb,Blanchet:2012vq,2012PhRvD..85b3004C,Calore:2014hna,Calore:2013yia}.
DM can give birth to $\gamma$ rays both directly (the so-called {\it prompt} emission) during the annihilation process or indirectly via the ICS of the electrons (or positrons) produced by the DM annihilation off the ambient light of the Interstellar Radiation Field (ISRF).
The $\gamma$-ray flux $dN_{\rm DM}/dE$ is described by \cite{1990NuPhB.346..129B,Bergstrom:1997fj,Bottino:2004qi,Calore:2013yia}
\begin{eqnarray}
\label{eq:dmflux}
  \frac{dN_{\rm DM}}{dE}(E) &=& \frac{1}{4 \pi } \frac{\langle\sigma v\rangle}{2 m_{\chi}^2} 
   \cdot \frac{1}{\Delta \Omega}\int_{\Delta \Omega} d\Omega  \int_{l.o.s.} d\lambda  \cdot \\	\nonumber 
&& \cdot \rho^{2}(r(\lambda, \psi)) f(E, r(\lambda, \psi)) \, ,
\end{eqnarray}
where  $m_{\chi}$ is the mass of the DM particle $\chi$ and \sigmav is the annihilation cross section times the relative velocity averaged over the DM velocity distribution. 
The last term  contains  the integral along the l.o.s. $\lambda$ of the squared DM density distribution $\rho(r)$ ($r$ being the galactocentric distance), 
where $\psi$ is the angle between the l.o.s.\,and the direction towards the 
Galactic center, defined as a function of the Galactic latitude $b$ and longitude $l$ 
($\cos\psi = \cos b\cos l$). When comparing with experimental data, an 
average over the telescope viewing solid angle $\Delta \Omega$ must be performed. 
The function  $f(E, r(\lambda, \psi))$ is the $\gamma$-ray energy spectrum:
\begin{eqnarray}
f(E, r(\lambda, \psi)) \equiv 
\left\{
\begin{array}{l}
\sum_i B_i \frac{d{\cal N}_i}{dE} (E)   \;\;\; {\rm  (prompt)}  \\
\int_{m_e}^{m_{\chi}} \sum_i B_i \frac{d{\cal N}_{i}}{dE_e} (E_e) \cdot \\ \cdot I_{IC} (E_e,E,r(\lambda, \psi)) dE_e   \;\;\; {\rm (ICS)} \end{array}
\right.
\end{eqnarray}
where $B_i$ is the branching ratio into the final state $i$,  and $d{\cal N}_i/dE$ and $d{\cal N}_i/dE_e$ are the photon and electron spectra per annihilation,
 which are summed over all annihilation channels. 
$I_{IC}(E_e,E,r(\lambda, \psi))$ is the halo function for the ICS radiative process and is given by \cite{Calore:2013yia,2009NuPhB.821..399C,Cirelli:2010xx}
\begin{eqnarray}
\label{eq:ichalo}
 I_{IC}(E_e,E,r(\lambda, \psi)) &=& 2 E \int_{m_e}^{E_e} \frac{\sum^3_{i=1} \mathcal{P}_i(E',E,r(\lambda, \psi))}{b(E',r(\lambda, \psi))} \cdot \nonumber \\
&& \cdot I (E',E_e,r(\lambda, \psi)) dE'  \,,
\end{eqnarray}
where $\mathcal{P}_i(E',E,r(\lambda, \psi))$ is the differential power emitted into photons due to ICS, 
and the sum runs over the different components of the ISRF:  the Cosmic Microwave Background (CMB), 
dust-rescattered light and starlight \cite{1970RvMP...42..237B,1994hea2.book.....L}. The term
$b(E',r(\lambda, \psi))$ takes into account the energy losses due to the ICS on the three ISRF photon fields and to synchrotron radiation  
\cite{1970RvMP...42..237B,1994hea2.book.....L}. 
Finally $I (E',E_e,r(\lambda, \psi))$ is the generalized halo function and is given by the Green function from a source with fixed energy $E_e$ to any energy $E'$. 
The halo functions contain all the astrophysics information and are independent of the particle physics model.
Neglecting the diffusion on the Galactic magnetic fields, 
 one would have $I (E',E_e,r(\lambda, \psi)) = 1$. However, we have included it adopting the MED model derived in \cite{Donato:2003xg}. 
Concerning the energy losses, we have used the same model as in \cite{2009NuPhB.821..399C,Cirelli:2010xx}.
\newline
In order to simplify the discussion, we do not consider any specific particle physics model.
For ease of presentation, we work at fixed branching ratio, set equal to 1 for any of the discussed annihilation channels.
The photon $d{\cal N}_i/dE$ and electron spectra $d{\cal N}_i/dE_e$ have been derived for DM annihilations into $e^{+}e^{-}$, $\mu^{+}\mu^{-}$, $\tau^{+}\tau^{-}$, $b\bar{b}$, $t\bar{t}$ and $W^+W^{-}$ 
channels using the Pythia Montecarlo code (version 8.162) \cite{Sjostrand:2007gs}.
We have considered an Einasto profile of DM with a local density of $\rho_{\odot}=0.4$ GeV/cm$^3$ \cite{2010A&A...523A..83S,2010JCAP...08..004C} and a distance of the Sun from the Galactic center of 8.33 kpc \cite{Gillessen:2008qv,2009ApJ...704.1704B,Ghez:2008ms}.
The analysis is performed for latitudes $ |b|>20^\circ$.
For all the details about the $\gamma$-ray flux due to DM annihilation in the halo of the MW we refer to \cite{Calore:2013yia}.

\subsection{Fits to the IGRB: Emission from extragalactic sources and DM annihilation}
\label{sect:DM}
\begin{figure*}
\centering
\includegraphics[width=\columnwidth]{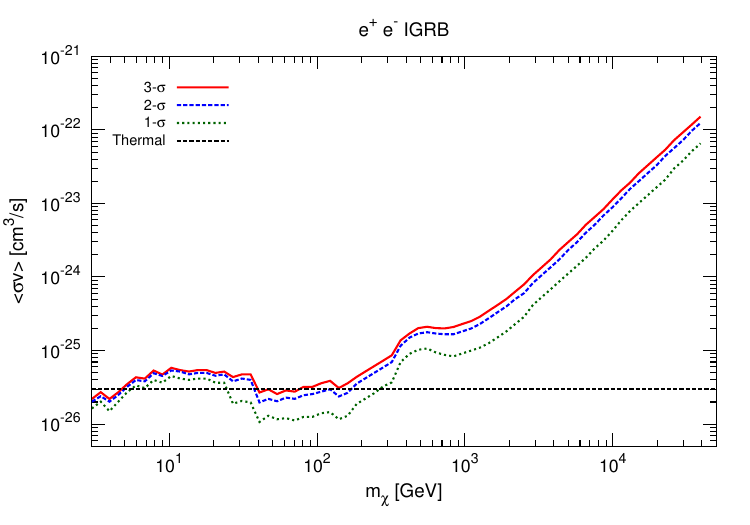}
\includegraphics[width=\columnwidth]{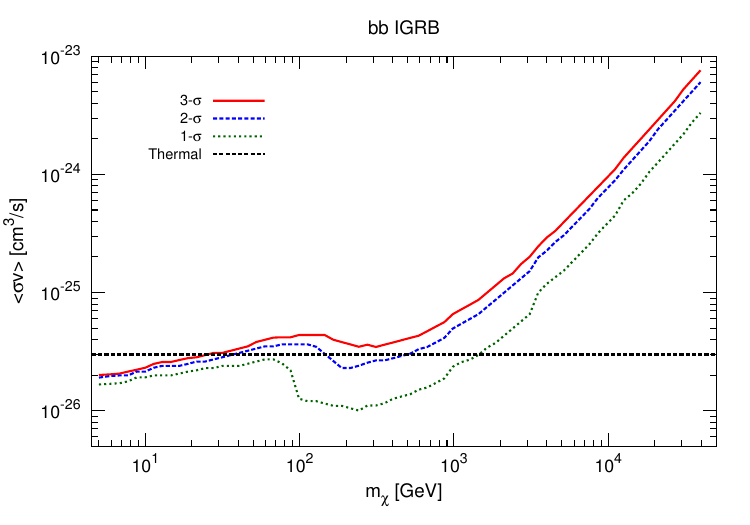}
\includegraphics[width=\columnwidth]{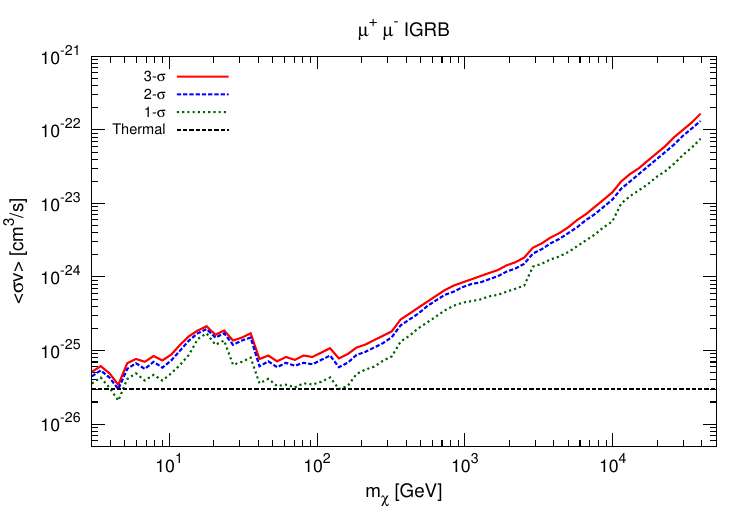}
\includegraphics[width=\columnwidth]{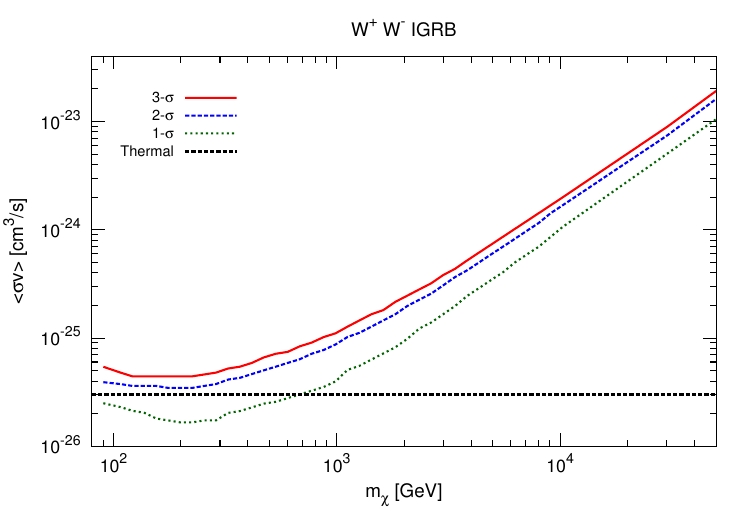}
\includegraphics[width=\columnwidth]{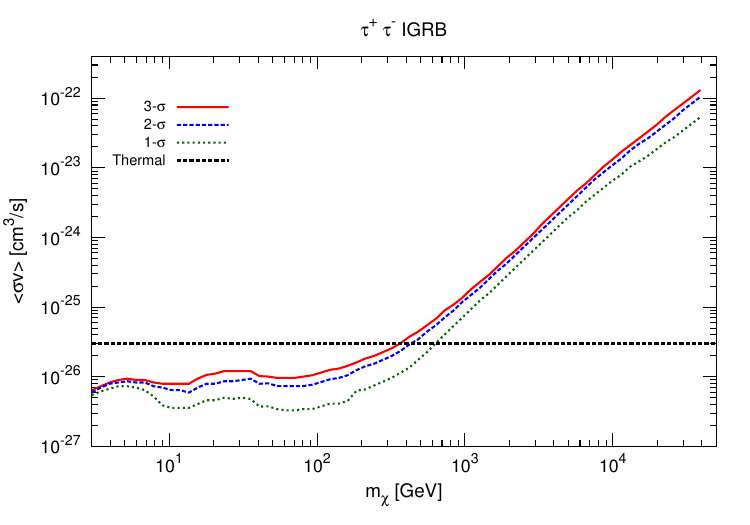}
\includegraphics[width=\columnwidth]{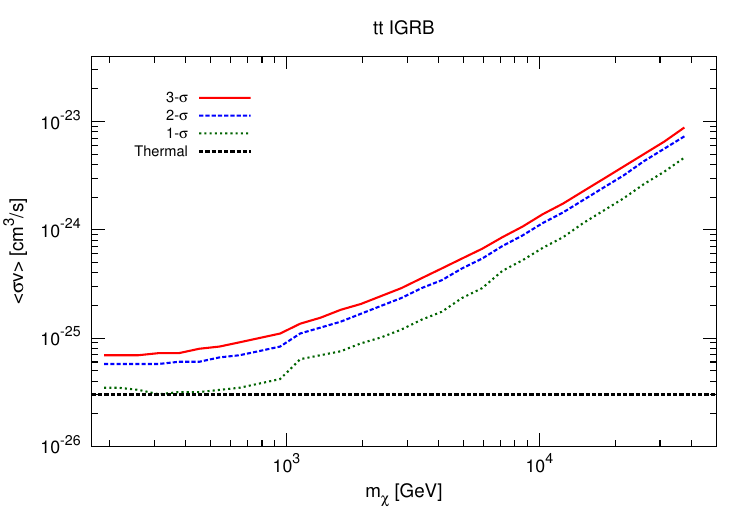}
\caption{Upper limits on the DM annihilation cross section \sigmav as a function of the DM mass $m_\chi$, at fixed annihilation channel into 
$\gamma$-rays (branching ratio=1). The limits are derived from a fit to the IGRB data, within Galactic foreground model A. 
Left  (right) top, middle and lower panels show the annihilation into $e^+e^-, \mu^+\mu^-, \tau^+\tau^-$ ($b\bar{b}, W^+W^-, t\bar{t}$), respectively.  
Solid, dashed and dotted curves correspond to 3-$\sigma$, 2-$\sigma$ and 1-$\sigma$ C.L.s, respectively. The dotted horizontal line
indicates the thermal relic  annihilation value $\langle\sigma v\rangle=3\cdot 10^{-26} {\rm cm}^3 / {\rm s}$.}
\label{fig:UL_IGRB}
\medskip
\end{figure*}

As a first analysis, we fit the IGRB data with the emission from the astrophysical sources discussed 
in Sec.~\ref{fit:bgd} and in the case their effective normalizations are left free ($M=5$), with the addition of 
Galactic DM  annihilating into $\gamma$ rays. 
We model the $\gamma$-ray emission from DM as described in Sec.~\ref{sect:dm},  fixing the WIMP DM mass $m_\chi$ and letting  \sigmav as a further free parameter and
consider DM annihilation in the halo of the Milky Way. 
We deduce for each DM mass the best-fit configuration (associated to the $\chi_{\rm{min}}^2$) given by the normalizations of the extragalactic sources emission and the DM annihilation cross section $\langle\sigma v\rangle$.
We then consider among all the configurations with a $\chi^2$ smaller than $\chi_{\rm{min}}^2 + \Delta \chi^2$ (where the $\Delta \chi^2$ is associate to one degree of freedom, namely $\langle\sigma v\rangle$) the configuration with the largest value of $\langle\sigma v\rangle$. This value of \sigmav is the DM upper limit for that DM mass.
Finally, following this method for a sampling of DM mass values, we derive the upper limit for each DM annihilation channel.
The results are shown in Fig.~\ref{fig:UL_IGRB} for different DM annihilation channels, and different 
confidence levels (C.L.s). The results have been obtained by fitting the IGRB data in the Galactic foreground 
scheme of model A. In the case annihilation proceeds via channel $e+e^-$, the cross section upper bound 
is about or slightly lower than the thermal relic  value $\langle\sigma v\rangle=3\cdot 10^{-26} {\rm cm}^3 / {\rm s}$ for $m_\chi \lsim 150 $ GeV, at 3-$\sigma$ C.L., 
while at $m_\chi=1$ TeV the bound is $2\cdot 10^{-25} {\rm cm}^3 / {\rm s}$ and for $m_\chi=20 $ TeV it raises to 
$5\cdot 10^{-23} {\rm cm}^3 / {\rm s}$. Similar results can be drawn for the $b\bar{b}$ channel. For masses lighter than 
30 GeV, the upper bounds on \sigmav are below the thermal relic value, and are slightly above up to $m_\chi \sim 500$ GeV. 
At 1 TeV our analysis excludes $\langle\sigma v\rangle \gsim 10^{-24} {\rm cm}^3 / {\rm s}$. 
The limits obtained for annihilation into $\tau^+\tau^-$ are quite stringent: for $m_\chi\lsim $ 100 GeV, $\langle\sigma v\rangle \lsim 10^{-26} {\rm cm}^3 / {\rm s}$
at 3-$\sigma$ C.L., while at $m_\chi \simeq1$ TeV the bound is around $10^{-25} {\rm cm}^3 / {\rm s}$. 
\begin{figure*}
\centering
\includegraphics[width=\columnwidth]{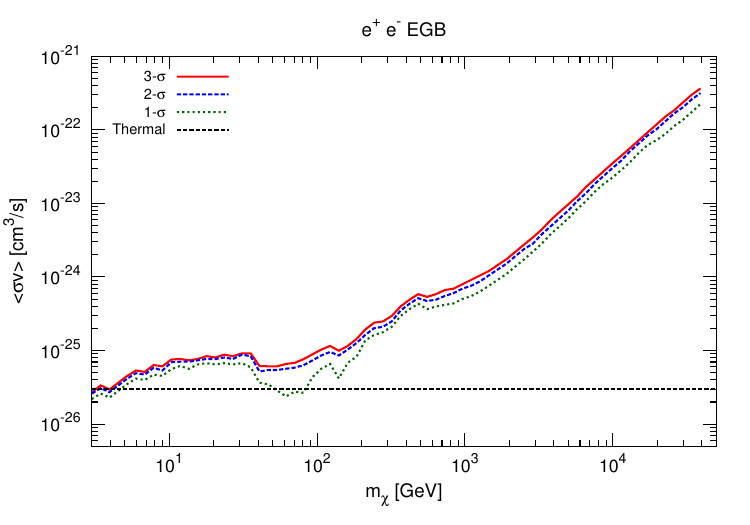}
\includegraphics[width=\columnwidth]{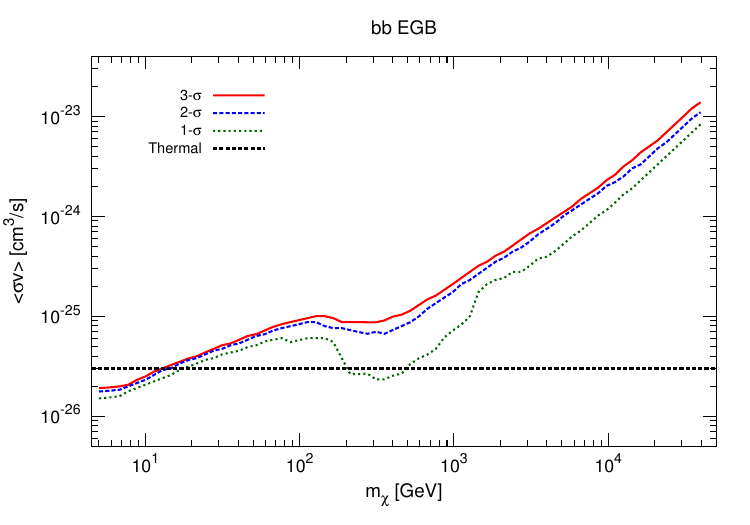}
\includegraphics[width=\columnwidth]{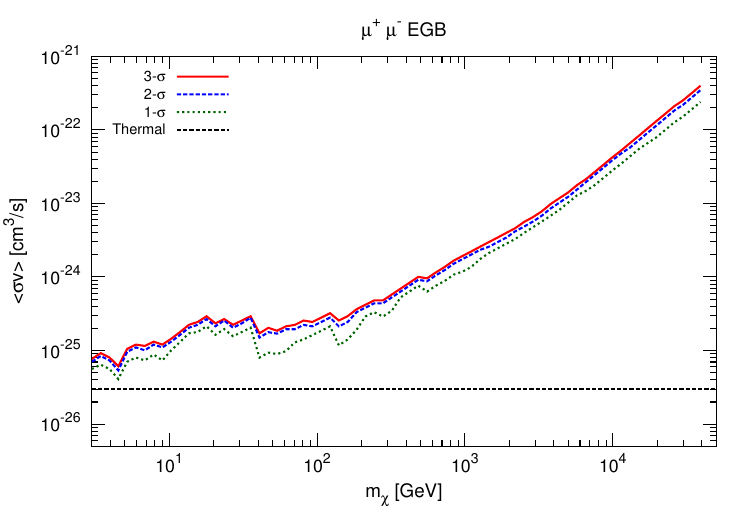}
\includegraphics[width=\columnwidth]{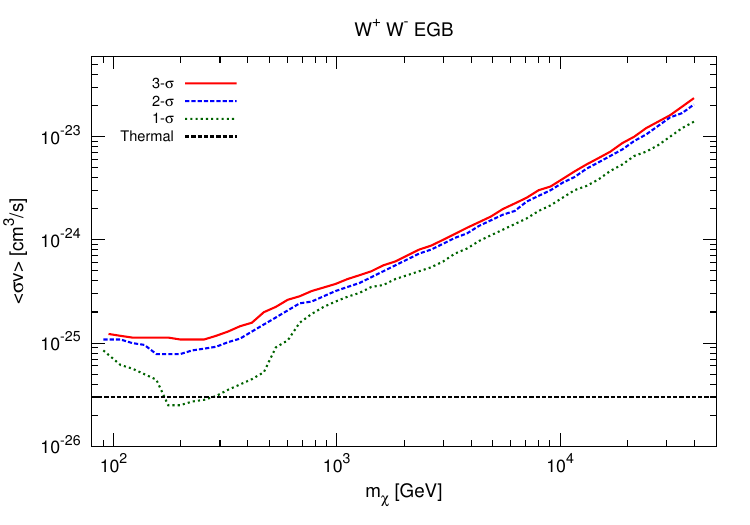}
\includegraphics[width=\columnwidth]{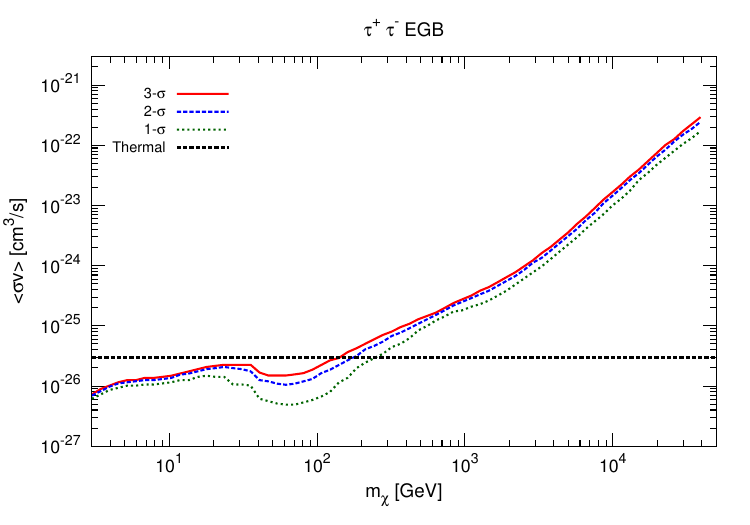}
\includegraphics[width=\columnwidth]{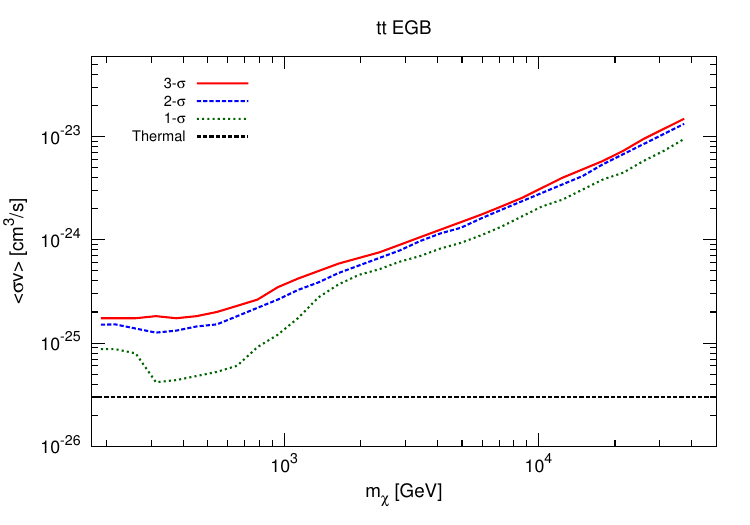}
\caption{Same as in Fig.~\ref{fig:UL_IGRB} but for the EGB instead of the IGRB data.}
\label{fig:UL_EGB}
\medskip
\end{figure*}

In Fig.~\ref{fig:UL_EGB} we plot the results obtained for the same analysis but using the data on the EGB, instead of the IGRB ones. 
As expected, the upper bounds obtained on the annihilation cross section into $\gamma$-rays are very similar to the ones 
shown in Fig.~\ref{fig:UL_IGRB}, and in general are only slightly looser.  
\begin{figure*}
\centering
\includegraphics[width=\columnwidth]{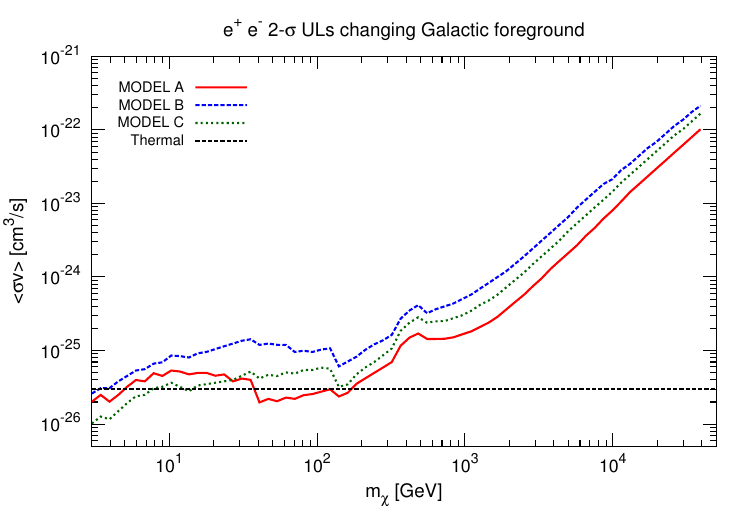}
\includegraphics[width=\columnwidth]{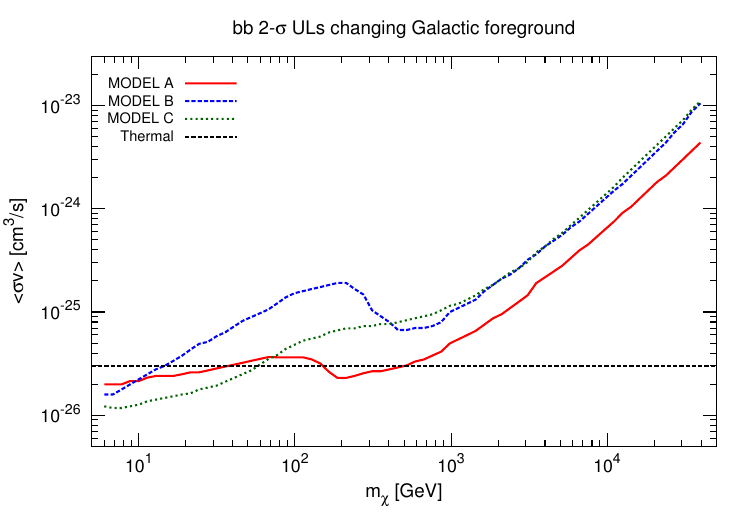}
\caption{Upper limits on the annihilation cross section into $\gamma$ rays (at 2-$\sigma$ C.L.) 
derived using the IGRB data obtained with three different Galactic foreground models (labeled A, B, and C as in \cite{igrb_2014}). Left and right panels refer to annihilation into $e^+e^-$ and $b\bar{b}$, respectively.}
\label{fig:gal_foreground}
\medskip
\end{figure*}

We also explore the relevance (if any) of the foreground Galactic model employed for obtaining the IGRB data and similarly for the EGB in the extraction of upper limits on the DM annihilation strength. 
The results are illustrated in Fig.~\ref{fig:gal_foreground} for the annihilation channels $e^+e^-$ and $b\bar{b}$, 
and for a 2-$\sigma$ C.L. The foreground Galactic models are labeled here A, B and C following  \cite{igrb_2014}. 
Indeed, the modeling of the Galactic emission turn out to have a significant 
role for the extraction of the upper limits on $\langle\sigma v\rangle$, which can vary up to a factor of ten. 
The IGRB obtained with Galactic model B always provides limits which are looser with respect to the use of model A or C. 
The reason is that the IGRB data within model B of the Galactic foreground models are systematically higher than for model A and C.
On average, the differences among the models are within a factor of two, but in the case of 
$b\bar{b}$ at about $m_\chi \simeq 200$ GeV they span one order of magnitude. This result is consistent  
with the IGRB shape and intensity derived in these three cases, as shown in Fig.~\ref{fig:backfit}. 

\begin{figure*}
\centering
\includegraphics[width=\columnwidth]{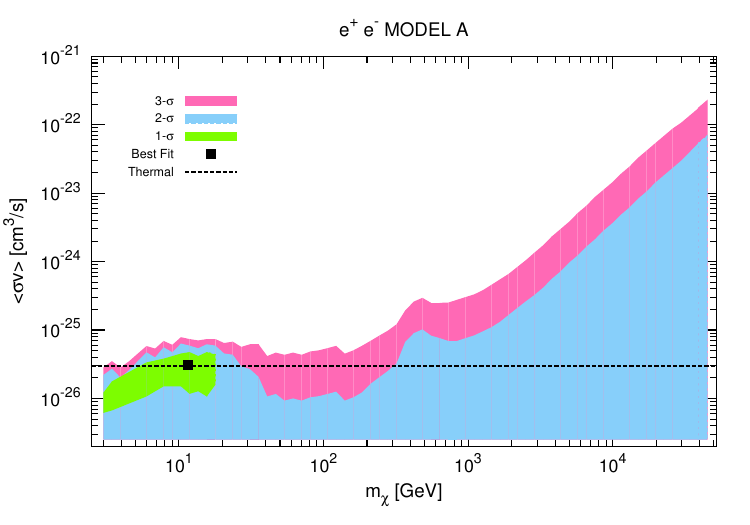}
\includegraphics[width=\columnwidth]{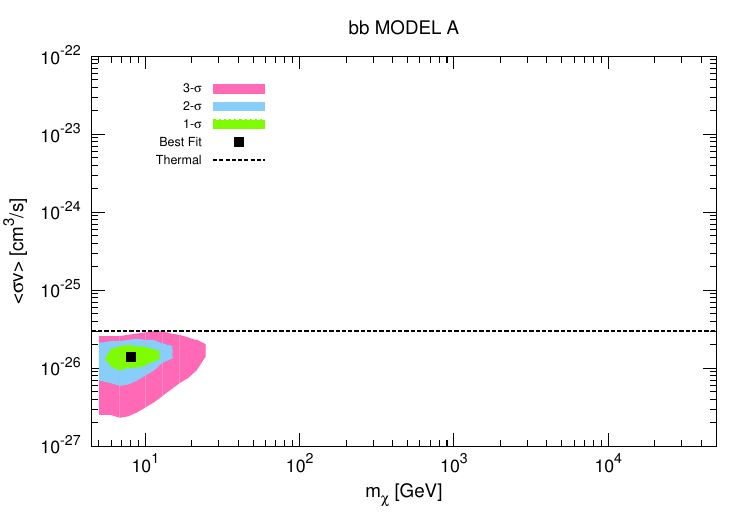}
\includegraphics[width=\columnwidth]{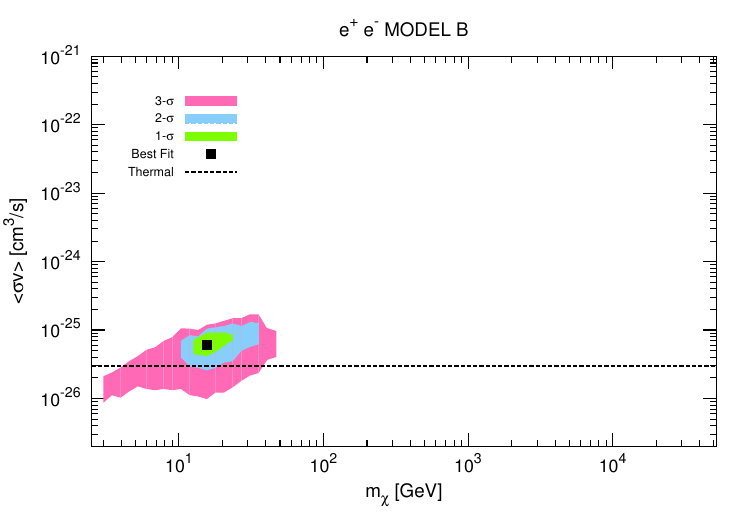}
\includegraphics[width=\columnwidth]{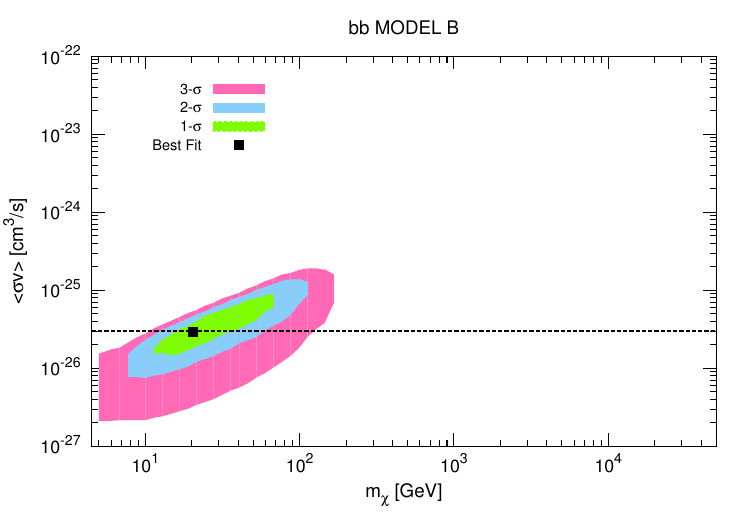}
\includegraphics[width=\columnwidth]{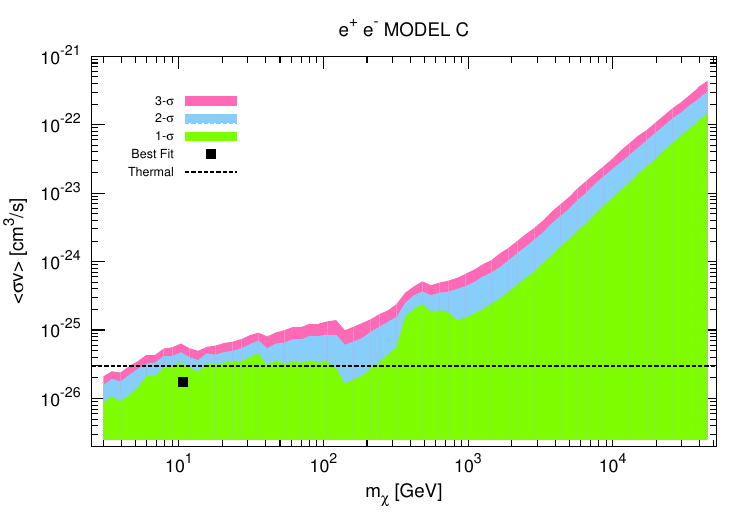}
\includegraphics[width=\columnwidth]{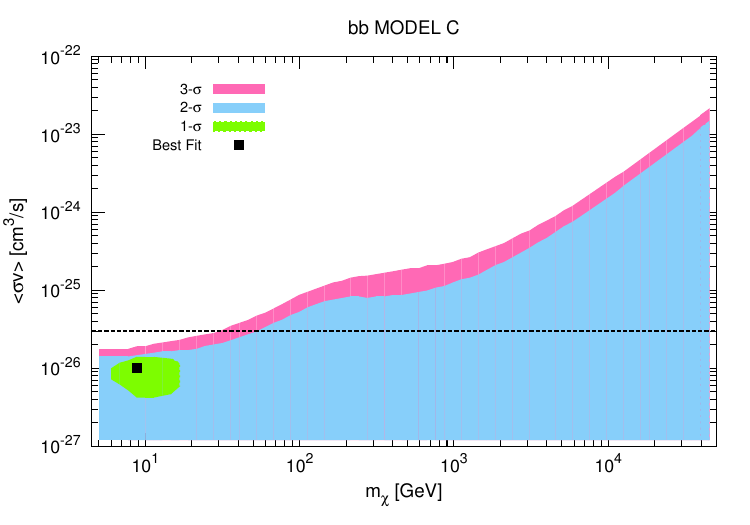}
\caption{Contour plots in the $\langle\sigma v\rangle$-$m_\chi$ plane obtained from the fit of the IGRB data with the astrophysical backgrounds and a DM component. The black dots identify the values for the best fit, and the (closed or open) pink, blue and green regions correspond to 3, 2 and 1-$\sigma$ C.L. Upper, medium and lower panels refer to model A, B and C of the IGRB, while left and right columns to  $e^+e^-$ and  $b\bar{b}$ DM annihilation channels, respectively. }
\label{fig:cp_igrb}
\medskip
\end{figure*}

As a second analysis, we fit the IGRB in terms of extragalactic sources and a Galactic DM component 
as done before, but  letting the WIMP DM mass $m_\chi$ and  \sigmav varying simultaneously. 
We end up with $M=7$, namely five free normalizations for the background and two DM parameters.
This procedure permits one not only to establish upper limits, but also to identify a DM configuration, if any, which can improve the fit to the IGRB data. 
Our results are plotted in Fig.~\ref{fig:cp_igrb}, for the IGRB data derived within the three Galactic foreground  models A, B and C and for the two representative DM annihilation channels $e^+e^-$ and  $b\bar{b}$. The fit has been performed varying the astrophysical diffuse emission from BL Lacs, FSRQs,  MAGN and SF (MW model) with their free overall normalizations, and with \sigmav and $m_\chi$ as free parameters for the DM sector.  
Depending on the IGRB data employed in the fit - namely on the foreground Galactic model - and on the required C.L., 
we obtain either closed regions or upper limits. In the case of model B, we obtain closed regions up to 3-$\sigma$ C.L. around 
$m_\chi \simeq$ 16 GeV and  \sigmav $\simeq 6\cdot 10^{-26}\rm{cm}^3/\rm{s}$ for $e^+e^-$, and $m_\chi \simeq$ 20 GeV and \sigmav $\simeq 10^{-26}\rm{cm}^3/\rm{s}$ for $b\bar{b}$. 
For the latter channel, we also find closed regions for data in model A at somewhat lower masses, while for model C the 1-$\sigma$ C.L. 
closed region opens up at already 2-$\sigma$ C.L., translating the results into upper limits. 

The details of these fits to the IGRB, and similarly to the EGB, are reported in Tabs. \ref{tab:fitdmbackigrb} and \ref{tab:fitdmbackegb}, respectively, where we extend 
the analysis also to the other leptonic annihilation channels. We also add the $\chi^2$ value for the best fit with the DM component, and the difference $\Delta \chi^2$ between 
this value and the $\chi^2$ obtained when only the astrophysical components ($i.e.$ no DM) are fitted. On general grounds, we can notice that the addition of a DM component 
improves the IGRB fit for models A and B, while it is almost irrelevant for IGRB model C. Depending on the annihilation channel, the best-fit DM mass ranges from few GeV up to 
20 GeV, while for $\langle\sigma v\rangle$  the fit chooses values close to the thermal one ($3\cdot 10^{-26}\rm{cm}^3/\rm{s}$), with the exception of the 
$\mu^+\mu^-$ channel, which requires significantly higher fluxes in order to improve the data fit. The numbers in the Table~\ref{tab:fitdmbackigrb} also 
show that the addition of a DM component does not require one to modify the normalization of the other astrophysical components with respect 
to their average values determined on independent theoretical grounds. That is to say,  a DM component can very well fit the IGRB data {\it together}
with the realistic emission from a number of unresolved extragalactic sources. 
Very similar comments hold for the fit to the EGB data reported in Table \ref{tab:fitdmbackigrb}. 

The results illustrated in Fig.~\ref{fig:cp_igrb} demonstrate indeed that a DM component $may$ improve the fit to the IGRB data with respect to 
the interpretation with only the diffuse emission from unresolved extragalactic populations. Nevertheless, this potential exotic signal may be 
easily misunderstood by different evaluations of the standard Galactic contribution that acts as a foreground for the IGRB derivation (see Appendix A in Reference 
\cite{igrb_2014} for details on the foreground models). Our results confirm how a deep knowledge of the Galactic $\gamma$-ray emission 
is demanded also for an unmistakable interpretation of the IGRB data with an additional DM component. 
We also notice that the results reported in Fig.~\ref{fig:cp_igrb} and in Tabs. \ref{tab:fitdmbackigrb} and \ref{tab:fitdmbackegb} may be considered in some tensions 
with upper limits derived from the AMS-02 positron fraction \cite{Bergstrom:2013jra} and the Pamela antiproton \cite{Fornengo:2013xda} data. 
However we remark that all these results, including ours, are at some extent  model dependent, {\it i.e.} 
on the assumed background from known sources, the propagation of charged particles in the Galaxy and in the heliosphere, the statistical data analysis, the DM cosmological and particle physics modeling.
\begin{figure*}
\centering
\includegraphics[width=\columnwidth]{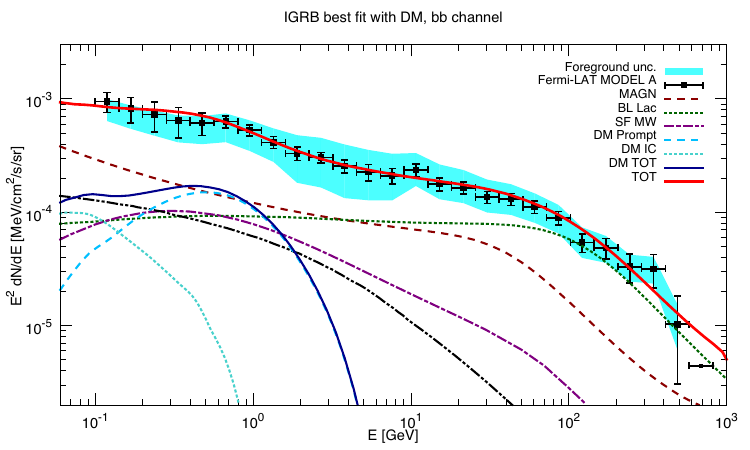}
\includegraphics[width=\columnwidth]{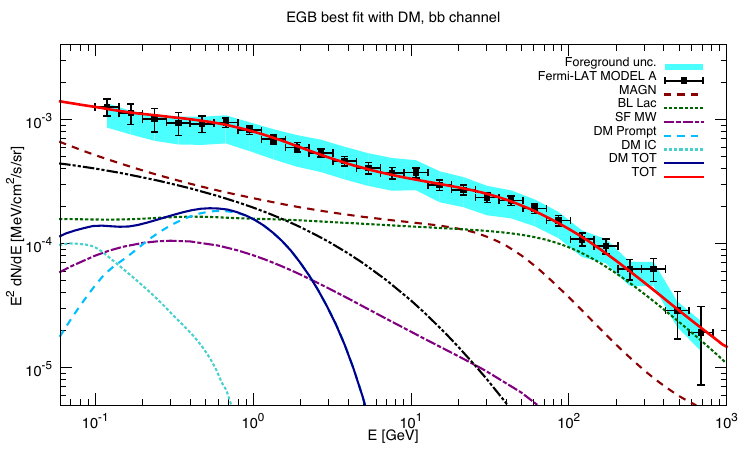}
\caption{Left (right) panel: Differential $\gamma$-ray flux for the unresolved (unresolved and resolved) BL Lac, FSRQ, MAGN, SF galaxy populations and the DM contribution
as fixed by the best fit to the IGRB  (EGB) data, model A (see Table~\ref{tab:fitdmbackigrb}). The DM annihilates through $b\bar{b}$ channel. Its flux is also split into the prompt and 
the ICS emission. The red solid line displays the sum  of all the contributions.  }
\label{fig:DM_bestfit}
\medskip
\end{figure*}
In Fig.~\ref{fig:DM_bestfit} we illustrate our results on the flux for one specific case in which the addition of a DM component improves 
significantly the fit to the IGRB. We plot the different contributions to the IGRB from the unresolved sources of the BL Lac, FSRQ, MAGN
and SF galaxy populations, plus the Galactic DM component as given by the best fit within $b\bar{b}$ annihilation channel to the IGRB data, 
model A (see Table \ref{tab:fitdmbackigrb}). In this specific case, the DM has $m_\chi$=8.2 GeV, and an annihilation cross section $\langle\sigma v\rangle
=1.4 \cdot 10^{-26}\rm{cm}^3/\rm{s}$. The figure shows clearly how well is the IGRB (and the EGB) represented by our model.

\begin{figure*}
\centering
\includegraphics[width=\columnwidth]{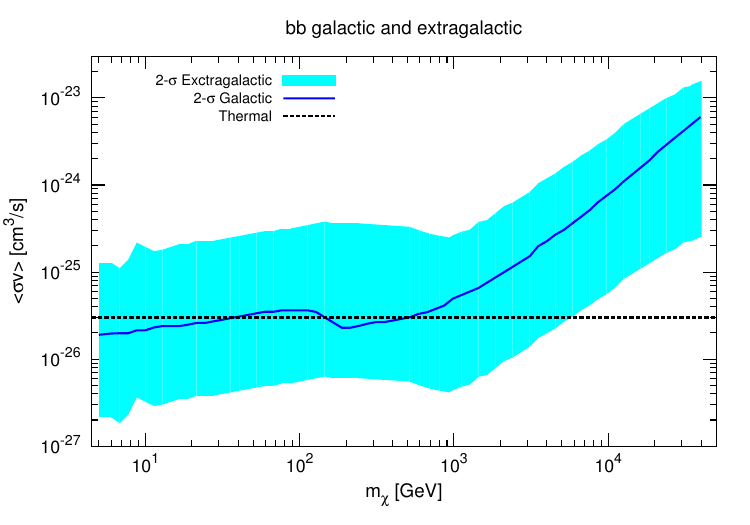}
\includegraphics[width=\columnwidth]{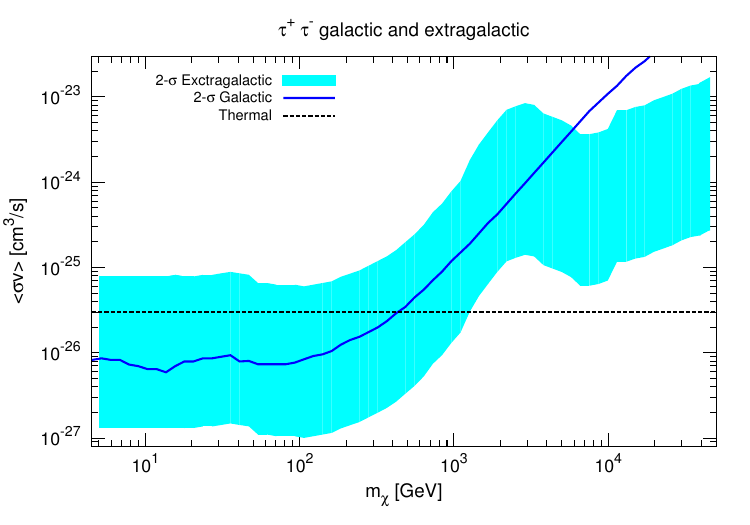}
\caption{Upper limits (at 2-$\sigma$ C.L.) on the DM annihilation cross section obtained from extragalactic DM (left and right panels are for 
$b\bar{b}$ and $\tau^+\tau^-$ annihilation channels, respectively). The uncertainties on 
the predicted flux translate into the cyan band on $\langle\sigma v\rangle$. For reference, we also draw the 
upper bound found from the Galactic DM halo (the same as in Fig.~\ref{fig:UL_IGRB}).  }
\label{fig:DM_extragal}
\medskip
\end{figure*}

Our results have been derived for ICS $\gamma$ rays produced by electrons and positrons propagating in the Galaxy according to the MED model \cite{Donato:2003xg}. This propagation model fits the B/C data \cite{Maurin:2002ua} and has recently been shown to correctly explain also the 
 low energy positrons and electrons AMS-02 measurements \cite{DiMauro:2014iia}, differently for the MIN, and partially for the MAX, reference models. 
 A change in the propagation model would affect negligibly our results 
 for the hadronic DM annihilation channels \cite{Cirelli:2010xx}. 
Concerning the leptonic channels, the MIN and MAX models would significantly change the ICS $\gamma$-ray emission. 
However, the addition of the prompt emission to the ICS one attenuates the differences in the total DM $\gamma$-ray flux due to propagation model, 
and consequently  the results on the annihilation cross section.

We notice that a DM Galactic signal (integrated above 20$^{\circ}$,
as done here) could not be simply bounded by the IGRB (a quantity
isotropic by definition) and in principle the constraints should be derived, for instance, including a spatial DM template in the fit and performing a morphological analysis.
A morphological analysis of {\it Fermi}-LAT data is beyond the
scope of this paper.
As remarked in~\cite{Ackermann:2015tah}, the smooth nonisotropic Galactic DM halo emission is
partially degenerate with other
Galactic diffuse templates, in particular with the inverse Compton one.   
However, Reference~\cite{Ackermann:2015tah} shows that this possible morphological confusion does not significantly alter 
the bounds on the DM annihilation  cross section. They find that for some $m_\chi\leq 250 $  GeV the results on the excluded $\langle\sigma v\rangle$
can get higher by up to 40\%, while being conservatively derived for masses higher than the TeV. The cross sections at the level of the sensitivity 
reach of the IGRB measurement are found not to significantly alter the results due the presence of a DM template in the IGRB. 
Given also that our analysis is carried on for three different Galactic diffuse emission models - which can be considered to bracket effectively the systematics in the 
misunderstandings of the diffuse emission - and for an intensity averaged on a huge sky region ($|b|>20^\circ$), the effect of a DM anisotropy from 
the smooth halo is not expected to alter significantly any of our results.  

Finally, we have estimated the contribution to the $\gamma$-ray sky from the extragalactic DM structures using directly the numerical tool obtained in \cite{Cirelli:2010xx}, 
where one can find a comprehensive study of the different assumptions that have to be considered for this $\gamma$-ray emission.
One  of the biggest uncertainties depends on the choice for the  halo mass function,  
which depends in particular on the halo concentration parameter. 
In \cite{Cirelli:2010xx} the halo mass function has been taken with the universal form introduced in \cite{1974ApJ...187..425P}.
Regarding the halo concentration, Reference~\cite{Cirelli:2010xx} parameterizes it within
two different models, named as "Macci\`o et al." and "power law" as considered in \cite{Maccio':2008xb}.
Moreover two typical values for the minimum halo mass can be taken into account: $10^{-6}$ or $10^{-9}\, M_{\odot}$  (see \cite{2009JCAP...06..014M,2009NJPh...11j5027B}) . 
The combination of these assumptions gives an uncertainty of about a factor of about 60 in the final (at redshift zero) $\gamma$-ray flux from extragalactic DM.
This uncertainty is definitely overwhelming with respect to the other possible variable ingredients, including  the extragalactic background light absorption modeling 
(see \cite{Cirelli:2010xx} for further details).
We have computed the flux including both prompt and ICS photons, choosing the `minimal UV' model for the intergalactic stellar light \cite{Cirelli:2010xx} 
(we have verified that the `maximal UV' option has negligible effects on our results). 
The upper bounds on $\langle\sigma v\rangle$  derived from extragalactic DM are shown in Fig.~\ref{fig:DM_extragal}.
The uncertainties on the predicted flux translate into the cyan band on the annihilation cross section, which spans almost 2 order of magnitude 
(as noticed in \cite{Cirelli:2010xx}, the computation is performed within a Navarro-Frenk-White halo profile, and the analysis of different halo density shapes would add a further uncertainty of roughly an order of magnitude). 
From Fig.~\ref{fig:DM_extragal}, we can notice that the bounds set from the extragalactic DM encompass the ones derived from the mere Galactic DM component. Given the huge uncertainty of the extragalactic halo modeling, it is not possible to set stronger bounds with respect to the ones obtained from the smooth Galactic halo. 
Additional uncertainties on the extragalactic DM component are due to the DM distribution at small scales and to the effect of baryons in DM simulations (see e.g. \cite{Sefusatti:2014vha,2012MNRAS.421L..87S}).

\begin{table*}
\scalebox{0.9}{
\begin{tabular}{|c|c|c|c|c|c|c|c|c|c|}
\hline
 IGRB (A)  &   BL Lac     &   FSRQ     &  MAGN    &  SF  & $m_\chi$ (GeV)  & $\langle\sigma v\rangle (10^{-26}\rm{cm}^3/\rm{s})$  &  $\chi^2$ &  $\tilde{\chi}^2$  &  $\Delta 
 \chi^2$   \\
\hline
$e$  &   $0.92\pm0.05$   &  $1.01\pm0.06$   &  $0.72\pm0.08$   & $1.02\pm0.18$   & $11.6 \pm 1.3$ & $3.1 \pm 1.3$ &  27.0  & 1.50 &  7.4   \\
$\mu$   &   $0.92\pm0.05$   &  $1.01\pm0.06$   &  $0.72\pm0.08$   & $1.06\pm0.18$   & $20.3 \pm 4.3$ & $15.8 \pm 5.6$ &  27.9 & 1.55  &  6.5 \\
$\tau$   &   $0.92\pm0.05$   &  $1.01\pm0.06$   &  $0.75\pm0.07$   & $1.00\pm0.09$   & $3.3 \pm 0.6$ & $0.45 \pm 0.13$ &  22.4  & 1.24  &  12.0 \\
$b$  &   $0.93\pm0.05$   &  $1.00\pm0.06$   &  $0.73\pm0.07$   & $0.99\pm0.09$   & $8.2 \pm 2.3$ & $1.4 \pm 0.3$ &  15.5  & 0.86  &  18.9  \\
\hline
\hline
 IGRB (B)  &   BL Lac   &   FSRQ   &  MAGN  &  SF  & $m_\chi$ (GeV)  & $\langle\sigma v\rangle (10^{-26}\rm{cm}^3/\rm{s})$  &  $\chi^2$ &  $\tilde{\chi}^2$  &  $\Delta \chi^2$   \\
\hline
$e$  &   $0.99\pm0.06$   &  $1.00\pm0.06$   &  $1.24\pm0.11$   & $1.03\pm0.18$   & $15.6 \pm 2.0$ & $6.0 \pm 1.6$ &  9.6  & 0.53  &  16.9   \\
$\mu$   &   $0.99\pm0.05$   &  $1.01\pm0.06$   &  $1.20\pm0.12$   & $1.07\pm0.17$   & $19.5 \pm 3.6$ & $31.6 \pm 7.9$ &  8.4 & 0.47  &  18.1 \\
$\tau$   &   $0.98\pm0.06$   &  $1.01\pm0.06$   &  $1.35\pm0.10$   & $1.08\pm0.18$   & $5.1 \pm 1.6$ & $0.86 \pm 0.28$ &  14.4  & 0.80  &  12.1 \\
$b$  &   $0.99\pm0.06$   &  $1.00\pm0.06$   &  $1.28\pm0.10$   & $1.04\pm0.18$   & $20 \pm 8$ & $2.8 \pm 1.0$ &  10.8  & 0.60 &  15.7   \\
\hline
\hline
 IGRB (C)  &   BL Lac    &   FSRQ   &  MAGN    &  SF  & $m_\chi$ (GeV)  & $\langle\sigma v\rangle (10^{-26}\rm{cm}^3/\rm{s})$  &  $\chi^2$  &  $\tilde{\chi}^2$ &  $\Delta \chi^2$   \\
\hline
$e$  &   $0.94\pm0.05$   &  $1.00\pm0.06$   &  $0.84\pm0.08$   & $1.00\pm0.18$   & $10.5 \pm 1.6$ & $1.74 \pm 0.42$ &  14.1  & 0.80  &  2.3   \\
$\mu$   &   $0.94\pm0.05$   &  $1.01\pm0.06$   &  $0.87\pm0.08$   & $1.05\pm0.17$   & $11.3 \pm 3.3$ & $15.2 \pm 3.9$ &  15.2 & 0.81  &  1.2 \\
$\tau$   &   $0.94\pm0.05$   &  $1.01\pm0.06$   &  $0.87\pm0.08$   & $1.01\pm0.17$   & $3.4 \pm 0.4$ & $0.48 \pm 0.16$ &  15.2  & 0.81  &  1.2 \\
$b$  &   $0.95\pm0.05$   &  $1.00\pm0.05$    &  $0.83\pm0.08$   & $0.99\pm0.09$   & $8.8 \pm 1.5$ & $1.00 \pm 0.3$ &  11.5  & 0.64  &  4.9  \\
\hline
\end{tabular}
}
\caption{Results for the best fit and 1-$\sigma$ error to the IGRB data (model A, B and C) with BL Lacs, FSRQs, MAGN and SF galaxies (MW model) together with  a DM 
annihilation contribution characterized by the particle mass $m_\chi$  and  annihilation cross section $\langle\sigma v\rangle$. 
The numbers relevant to the BL Lac, FSRQ, MAGN and SF  columns refer to the best fit for the normalizations of the different 
contributions, evaluated with respect to their average theoretical values (see Sec.~\ref{fit:bgd} for details). 
The last three columns refer to the best fit  $\chi^2$, $\tilde{\chi}^2$ and to its difference $\Delta \chi^2$ with the best fit  $\chi^2$ obtained without the DM component, respectively. 
}
\label{tab:fitdmbackigrb}
\end{table*}
\begin{table*}
\scalebox{0.9}{
\begin{tabular}{|c|c|c|c|c|c|c|c|c|c|}
\hline
 EGB (A)  &   BL Lac     &   FSRQ     &  MAGN    &  SF  & $m_\chi$ (GeV)  & $\langle\sigma v\rangle (10^{-26}\rm{cm}^3/\rm{s})$  &  $\chi^2$ &  $\tilde{\chi}^2$  &  $\Delta 
 \chi^2$   \\
\hline
$e$  &   $1.03\pm0.05$   &  $1.01\pm0.06$   &  $1.15\pm0.12$      & $1.03\pm0.18$   & $15.0 \pm 3.2$ & $4.7 \pm 1.5$ &  9.6  & 0.51  &  10.4   \\
$\mu$   &   $1.02\pm0.05$  &  $1.03\pm0.06$   &  $1.16\pm0.13$   & $1.07\pm0.18$   & $20.3 \pm 3.7$ & $19.0 \pm 4.2$ &  11.8  & 0.62  &  8.2 \\
$\tau$   &   $1.02\pm0.05$   &  $1.02\pm0.06$   &  $1.24\pm0.11$   & $1.05\pm0.18$   & $5.2 \pm 0.9$ & $0.69 \pm 0.21$ &  11.7  & 0.62  &  8.3 \\
$b$  &   $1.03\pm0.05$   &  $1.00\pm0.06$  &  $1.21\pm0.12$   & $1.01\pm0.18$   & $10.6 \pm 3.1$ & $1.6 \pm 0.4$ &  8.3  & 0.44  &  11.7  \\
\hline
\hline
 EGB (B) &     BL Lac     &   FSRQ     &  MAGN    &  SF  & $m_\chi$ (GeV)  & $\langle\sigma v\rangle (10^{-26}\rm{cm}^3/\rm{s})$  &  $\chi^2$ &  $\tilde{\chi}^2$ &  $\Delta 
 \chi^2$    \\
\hline
$e$  &   $1.07\pm0.05$   &  $1.00\pm0.04$   &  $1.70\pm0.13$      & $1.02\pm0.18$   & $16.4 \pm 2.3$ & $7.8 \pm 1.8$ &  13.6  & 0.72  &  19.4   \\
$\mu$   &   $1.06\pm0.05$  &  $1.03\pm0.04$   &  $1.79\pm0.09$   & $1.12\pm0.14$   & $16.1 \pm 3.0$ & $20.3 \pm 4.1$ &  21.5  & 1.13  &  11.5 \\
$\tau$   &   $1.05\pm0.05$   &  $1.01\pm0.06$   &  $1.85\pm0.12$   & $1.07\pm0.18$   & $6.1 \pm 1.3$ & $1.13 \pm 0.35$ &  19.5  & 1.03  &  13.5 \\
$b$  &   $1.06\pm0.05$   &  $1.00\pm0.06$  &  $1.76\pm0.13$   & $1.04\pm0.18$   & $28.3 \pm 9.2$ & $5.0 \pm 1.4$ &  16.1  & 0.85  &  16.9  \\
\hline
\hline
 EGB (C) &   BL Lac     &   FSRQ     &  MAGN    &  SF  & $m_\chi$ (GeV)  & $\langle\sigma v\rangle(10^{-26}\rm{cm}^3/\rm{s})$  &  $\chi^2$ &  $\tilde{\chi}^2$  &  $\Delta 
 \chi^2$   \\
\hline
$e$  &   $1.04\pm0.05$   &  $1.00\pm0.05$   &  $1.28\pm0.13$      & $1.00\pm0.08$   & $15.0 \pm 3.2$ & $2.4 \pm 1.2$ &  8.4  & 0.44   &  4.2   \\
$\mu$   &   $1.04\pm0.05$  &  $1.00\pm0.04$   &  $1.27\pm0.12$   & $1.02\pm0.14$   & $19.5 \pm 2.8$ & $12.0 \pm 4.6$ &  8.3  & 0.44  &  4.3 \\
$\tau$   &   $1.03\pm0.05$   &  $1.02\pm0.06$   &  $1.33\pm0.12$   & $1.06\pm0.17$   & $6.5 \pm 1.7$ & $1.29 \pm 0.47$ &  9.7  & 0.51  &  2.9 \\
$b$  &   $1.04\pm0.05$   &  $1.00\pm0.04$  &  $1.32\pm0.12$   & $1.01\pm0.18$   & $23.9 \pm 6.1$ & $1.08 \pm 0.3$ &  9.3  & 0.49  &  3.3  \\
\hline
\end{tabular}
}
\caption{The same as in Table \ref{tab:fitdmbackigrb} but for the EGB data.}
\label{tab:fitdmbackegb}
\end{table*}

The results shown in Fig.~\ref{fig:UL_EGB} improve the upper bounds on the \sigmav by a factor of $\sim$3 at $m_{\chi} \sim 10$ GeV and a factor of at least 30 at 
$m_{\chi} \sim 10$ TeV in the so-called 'best-fit' scenario, while being comparable with the 'optimist 3s' model. 
Our limits also improve significantly the {\it Fermi} analysis for a Galactic halo of DM \cite{Ackermann:2012rg} both in the absence or presence of background modeling. 
At low DM masses, our results are comparable with the analysis of 25 dwarf spheroidal galaxies \cite{Ackermann:2013yva}, 
while they are lower  by about one order of magnitude at $m_{\chi} \sim 10$ TeV. 
Limits on \sigmav from dwarf spheroidal galaxies have also been derived by the H.E.S.S. and MAGIC  Collaborations in \cite{Aleksic:2013xea,Abramowski:2014tra}.
The analysis of these imaging atmospheric Cherenkov telescope data are optimized at energies larger than about 1 TeV. We notice that for 
$m_\chi\simeq$ 10 TeV the upper bounds to \sigmav found in \cite{Abramowski:2014tra} are of the same entity as ours for the leptonic channels, 
while for hadronic channels  they are about one order of magnitude weaker.
A similar discussion holds for the comparison with  bounds from radio emission 
\cite{2012JCAP...01..005F}. 
Finally, our bounds are significantly stronger than those obtained from galaxy cluster \cite{2010JCAP...05..025A} and CMB observations \cite{2011PhRvD..84b7302G}.

\section{Conclusions}
In this paper we have performed a detailed analysis of the recent data on the IGRB provided by 
the {\it Fermi}-LAT Collaboration, based on the analysis of 50 months of sky-survey observations  \cite{igrb_2014}. 
The data we work with refer to high ($|b| > 20^{\circ}$) latitudes and to the energy range 
from 100 MeV to 820 GeV. The experimental results are obtained for different modelings of the Galactic foreground, and we refer here 
to the models A, B and C as defined in \cite{igrb_2014}.
\\
Our first attempt is the interpretation of the  {\it Fermi}-LAT IGRB data in terms of the $\gamma$-ray unresolved emission from different extragalactic populations.
We find very good fits to the experimental IGRB data, obtained  with the theoretical predictions for the emission from BL Lacs, FSRQs, MAGN and SF galaxies. 
The flux from each component is let varying within the theoretical uncertainty band fixed from independent phenomenological analysis. 
We find that the IGRB (and very similarly also the EGB) is well fitted by the diffuse emission of AGN and SF galaxies. The best fit 
to the IGRB data is obtained for BL Lacs, FSRQs, MAGN and SF galaxy models extremely close to their most reliable theoretical predictions. 
Furthermore, we show that the choice for the foreground model employed in the data derivation is quite relevant in  the fits to the IGRB
data. In particular, we find that our models  provides always better fits to  IGRB data derived within model C. 
Moreover, the fit is always better when MW modeling of SF galaxies is chosen with respect to the PL one. 
\\
In addition to the emission from extragalactic sources, we probe a possible emission coming from the annihilation of 
WIMP DM in the halo of our Galaxy. 
As a first analysis, we set upper bounds on the DM annihilation cross section by the combined contribution from 
the above mentioned extragalactic source populations and annihilating DM. 
We set stringent limits on $\langle\sigma v\rangle$, which are about the thermal relic value 
for a wide range of DM masses. We also show how the data for different Galactic foreground models can change the upper 
limits on $\langle\sigma v\rangle$  by a significant factor. 
\\
As a final analysis, we seek for  a DM configuration, if any, which improves the fit to the IGRB data with respect to considering 
only the emission from extragalactic source populations. Depending on the IGRB data employed in the fit - namely on the foreground Galactic model - and on the required C.L., 
we obtain whether upper limits or closed regions in the  $\langle\sigma v\rangle-m_\chi$ parameter space. 
 On general grounds, we find  that the addition of a DM component 
improves the IGRB fit for models A and B, while it is almost irrelevant for IGRB model C. According to the annihilation channels, the best fit DM mass ranges from few GeV up to 
20 GeV, while for $\langle\sigma v\rangle$  the fit chooses values close to the thermal one. 
We also quantitatively  show that the addition of a DM component does not require to modify the normalizations of the other astrophysical components with respect 
to their average values determined on independent theoretical grounds. That is to say that a DM component can very well fit the IGRB data {\it together}
with the realistic emission from a number of unresolved extragalactic sources. 
\\
The results presented in this manuscript demonstrate how crucial the use of the IGRB is becoming in the study of the extragalactic sky and 
of DM searches. However, they also reinforce the strong need for a better understanding of the Galactic emission. A mandatory path to this goal $-$ 
already undertaken by several research groups $-$ 
includes a number of new data on the key players in the interactions leading to $\gamma$ rays (interstellar gas, magnetic fields, cosmic ray fluxes) 
and huge modeling efforts. 

\acknowledgments

M.D.M. thanks Andrea Vittino, Marco Ajello, Michael Gustafsson, Andrea Albert, Miguel \'A. S\'anchez-Conde, Anna Franckowiak and Alessandro Cuoco for useful discussions and comments.
We acknowledge Pasquale Dario Serpico and Francesca Calore for useful comments and a careful reading of the manuscript. 
This work is supported by the research grant {\sl TAsP (Theoretical Astroparticle Physics)}
funded by the Istituto Nazionale di Fisica Nucleare (INFN), by the  {\sl Strategic Research Grant: Origin and Detection of Galactic and Extragalactic Cosmic Rays} funded by Torino University and Compagnia di San Paolo. 
This work is also supported by the research grant {\sl Theoretical 
Astroparticle Physics} number 2012CPPYP7 under the program PRIN 2012  
funded by the Ministero dell'Istruzione, Universit\`a e della Ricerca 
(MIUR).
Finally, at LAPTh this activity was supported by the Labex grant ENIGMASS.

\bibliography{paper_referee_2}

\begin{thebibliography}{99}
\expandafter\ifx\csname natexlab\endcsname\relax\def\natexlab#1{#1}\fi
\expandafter\ifx\csname bibnamefont\endcsname\relax
  \def\bibnamefont#1{#1}\fi
\expandafter\ifx\csname bibfnamefont\endcsname\relax
  \def\bibfnamefont#1{#1}\fi
\expandafter\ifx\csname citenamefont\endcsname\relax
  \def\citenamefont#1{#1}\fi
\expandafter\ifx\csname url\endcsname\relax
  \def\url#1{\texttt{#1}}\fi
\expandafter\ifx\csname urlprefix\endcsname\relax\def\urlprefix{URL }\fi
\providecommand{\bibinfo}[2]{#2}
\providecommand{\eprint}[2][]{\url{#2}}

\bibitem[{\citenamefont{{Kraushaar} et~al.}(1972)\citenamefont{{Kraushaar},
  {Clark}, {Garmire}, {Borken}, {Higbie}, {Leong}, and
  {Thorsos}}}]{1972ApJ...177..341K}
\bibinfo{author}{\bibfnamefont{W.~L.} \bibnamefont{{Kraushaar}}},
  \bibinfo{author}{\bibfnamefont{G.~W.} \bibnamefont{{Clark}}},
  \bibinfo{author}{\bibfnamefont{G.~P.} \bibnamefont{{Garmire}}},
  \bibinfo{author}{\bibfnamefont{R.}~\bibnamefont{{Borken}}},
  \bibinfo{author}{\bibfnamefont{P.}~\bibnamefont{{Higbie}}},
  \bibinfo{author}{\bibfnamefont{V.}~\bibnamefont{{Leong}}}, \bibnamefont{and}
  \bibinfo{author}{\bibfnamefont{T.}~\bibnamefont{{Thorsos}}},
  \bibinfo{journal}{\apj} \textbf{\bibinfo{volume}{177}}, \bibinfo{pages}{341}
  (\bibinfo{year}{1972}).

\bibitem[{\citenamefont{{Fichtel} et~al.}(1975)\citenamefont{{Fichtel},
  {Hartman}, {Kniffen}, {Thompson}, {Ogelman}, {Ozel}, {Tumer}, and
  {Bignami}}}]{1975ApJ...198..163F}
\bibinfo{author}{\bibfnamefont{C.~E.} \bibnamefont{{Fichtel}}},
  \bibinfo{author}{\bibfnamefont{R.~C.} \bibnamefont{{Hartman}}},
  \bibinfo{author}{\bibfnamefont{D.~A.} \bibnamefont{{Kniffen}}},
  \bibinfo{author}{\bibfnamefont{D.~J.} \bibnamefont{{Thompson}}},
  \bibinfo{author}{\bibfnamefont{H.}~\bibnamefont{{Ogelman}}},
  \bibinfo{author}{\bibfnamefont{M.~E.} \bibnamefont{{Ozel}}},
  \bibinfo{author}{\bibfnamefont{T.}~\bibnamefont{{Tumer}}}, \bibnamefont{and}
  \bibinfo{author}{\bibfnamefont{G.~F.} \bibnamefont{{Bignami}}},
  \bibinfo{journal}{\apj} \textbf{\bibinfo{volume}{198}}, \bibinfo{pages}{163}
  (\bibinfo{year}{1975}).

\bibitem[{\citenamefont{{Sreekumar} et~al.}(1998)\citenamefont{{Sreekumar},
  {Bertsch}, {Dingus}, {Esposito}, {Fichtel}, {Hartman}, {Hunter}, {Kanbach},
  {Kniffen}, {Lin} et~al.}}]{1998ApJ...494..523S}
\bibinfo{author}{\bibfnamefont{P.}~\bibnamefont{{Sreekumar}}},
  \bibinfo{author}{\bibfnamefont{D.~L.} \bibnamefont{{Bertsch}}},
  \bibinfo{author}{\bibfnamefont{B.~L.} \bibnamefont{{Dingus}}},
  \bibinfo{author}{\bibfnamefont{J.~A.} \bibnamefont{{Esposito}}},
  \bibinfo{author}{\bibfnamefont{C.~E.} \bibnamefont{{Fichtel}}},
  \bibinfo{author}{\bibfnamefont{R.~C.} \bibnamefont{{Hartman}}},
  \bibinfo{author}{\bibfnamefont{S.~D.} \bibnamefont{{Hunter}}},
  \bibinfo{author}{\bibfnamefont{G.}~\bibnamefont{{Kanbach}}},
  \bibinfo{author}{\bibfnamefont{D.~A.} \bibnamefont{{Kniffen}}},
  \bibinfo{author}{\bibfnamefont{Y.~C.} \bibnamefont{{Lin}}},
  \bibnamefont{et~al.}, \bibinfo{journal}{\apj} \textbf{\bibinfo{volume}{494}},
  \bibinfo{pages}{523} (\bibinfo{year}{1998}), \eprint{astro-ph/9709257}.

\bibitem[{\citenamefont{{Abdo} et~al.}(2010{\natexlab{a}})\citenamefont{{Abdo},
  {Ackermann}, {Ajello} et~al.}}]{2010PhRvL.104j1101A}
\bibinfo{author}{\bibfnamefont{A.~A.} \bibnamefont{{Abdo}}},
  \bibinfo{author}{\bibfnamefont{M.}~\bibnamefont{{Ackermann}}},
  \bibinfo{author}{\bibfnamefont{M.}~\bibnamefont{{Ajello}}},
  \bibnamefont{et~al.}, \bibinfo{journal}{Phys. Rev. Lett.}
  \textbf{\bibinfo{volume}{104}}, \bibinfo{eid}{101101}
  (\bibinfo{year}{2010}{\natexlab{a}}), \eprint{1002.3603}.

\bibitem[{\citenamefont{Ackermann et~al.}(2014{\natexlab{a}})}]{igrb_2014}
\bibinfo{author}{\bibfnamefont{M.}~\bibnamefont{Ackermann}}
  \bibnamefont{et~al.}, \bibinfo{journal}{\apj} \textbf{\bibinfo{volume}{in
  press}} (\bibinfo{year}{2014}{\natexlab{a}}), \eprint{1410.3696}.

\bibitem[{\citenamefont{{Chiang} and {Mukherjee}}(1998)}]{1998ApJ...496..752C}
\bibinfo{author}{\bibfnamefont{J.}~\bibnamefont{{Chiang}}} \bibnamefont{and}
  \bibinfo{author}{\bibfnamefont{R.}~\bibnamefont{{Mukherjee}}},
  \bibinfo{journal}{\apj} \textbf{\bibinfo{volume}{496}}, \bibinfo{pages}{752}
  (\bibinfo{year}{1998}).

\bibitem[{\citenamefont{{M{\"u}cke} and {Pohl}}(2000)}]{2000MNRAS.312..177M}
\bibinfo{author}{\bibfnamefont{A.}~\bibnamefont{{M{\"u}cke}}} \bibnamefont{and}
  \bibinfo{author}{\bibfnamefont{M.}~\bibnamefont{{Pohl}}},
  \bibinfo{journal}{\mnras} \textbf{\bibinfo{volume}{312}},
  \bibinfo{pages}{177} (\bibinfo{year}{2000}).

\bibitem[{\citenamefont{{Narumoto} and {Totani}}(2006)}]{narumoto2006}
\bibinfo{author}{\bibfnamefont{T.}~\bibnamefont{{Narumoto}}} \bibnamefont{and}
  \bibinfo{author}{\bibfnamefont{T.}~\bibnamefont{{Totani}}},
  \bibinfo{journal}{\apj} \textbf{\bibinfo{volume}{643}}, \bibinfo{pages}{81}
  (\bibinfo{year}{2006}), \eprint{astro-ph/0602178}.

\bibitem[{\citenamefont{{Dermer}}(2007)}]{2007ApJ...659..958D}
\bibinfo{author}{\bibfnamefont{C.~D.} \bibnamefont{{Dermer}}},
  \bibinfo{journal}{\apj} \textbf{\bibinfo{volume}{659}}, \bibinfo{pages}{958}
  (\bibinfo{year}{2007}), \eprint{astro-ph/0605402}.

\bibitem[{\citenamefont{Kneiske and Mannheim}(2007)}]{Kneiske:2007jq}
\bibinfo{author}{\bibfnamefont{T.~M.} \bibnamefont{Kneiske}} \bibnamefont{and}
  \bibinfo{author}{\bibfnamefont{K.}~\bibnamefont{Mannheim}},
  \bibinfo{journal}{Astron.Astrophys.} \textbf{\bibinfo{volume}{479}},
  \bibinfo{pages}{41} (\bibinfo{year}{2007}).

\bibitem[{\citenamefont{{Abazajian} et~al.}(2011)\citenamefont{{Abazajian},
  {Blanchet}, and {Harding}}}]{2011PhRvD..84j3007A}
\bibinfo{author}{\bibfnamefont{K.~N.} \bibnamefont{{Abazajian}}},
  \bibinfo{author}{\bibfnamefont{S.}~\bibnamefont{{Blanchet}}},
  \bibnamefont{and} \bibinfo{author}{\bibfnamefont{J.~P.}
  \bibnamefont{{Harding}}}, \bibinfo{journal}{\prd}
  \textbf{\bibinfo{volume}{84}}, \bibinfo{eid}{103007} (\bibinfo{year}{2011}),
  \eprint{1012.1247}.

\bibitem[{\citenamefont{{Inoue} and {Totani}}(2009)}]{2009ApJ...702..523I}
\bibinfo{author}{\bibfnamefont{Y.}~\bibnamefont{{Inoue}}} \bibnamefont{and}
  \bibinfo{author}{\bibfnamefont{T.}~\bibnamefont{{Totani}}},
  \bibinfo{journal}{\apj} \textbf{\bibinfo{volume}{702}}, \bibinfo{eid}{523}
  (\bibinfo{year}{2009}), \eprint{0810.3580}.

\bibitem[{\citenamefont{{Stecker} and {Salamon}}(1996)}]{1996ApJ...464..600S}
\bibinfo{author}{\bibfnamefont{F.~W.} \bibnamefont{{Stecker}}}
  \bibnamefont{and} \bibinfo{author}{\bibfnamefont{M.~H.}
  \bibnamefont{{Salamon}}}, \bibinfo{journal}{\apj}
  \textbf{\bibinfo{volume}{464}}, \bibinfo{pages}{600} (\bibinfo{year}{1996}),
  \eprint{astro-ph/9601120}.

\bibitem[{\citenamefont{{Stecker} and {Venters}}(2011)}]{2011ApJ...736...40S}
\bibinfo{author}{\bibfnamefont{F.~W.} \bibnamefont{{Stecker}}}
  \bibnamefont{and} \bibinfo{author}{\bibfnamefont{T.~M.}
  \bibnamefont{{Venters}}}, \bibinfo{journal}{\apj}
  \textbf{\bibinfo{volume}{736}}, \bibinfo{pages}{40} (\bibinfo{year}{2011}),
  \eprint{1012.3678}.

\bibitem[{\citenamefont{Neronov and Semikoz}(2012)}]{Neronov:2011kg}
\bibinfo{author}{\bibfnamefont{A.}~\bibnamefont{Neronov}} \bibnamefont{and}
  \bibinfo{author}{\bibfnamefont{D.}~\bibnamefont{Semikoz}},
  \bibinfo{journal}{\apj} \textbf{\bibinfo{volume}{757}}, \bibinfo{pages}{61}
  (\bibinfo{year}{2012}), \eprint{1103.3484}.

\bibitem[{\citenamefont{Ajello et~al.}(2014{\natexlab{a}})\citenamefont{Ajello,
  Romani, Gasparrini, Shaw, Bolmer et~al.}}]{Ajello:2013lka}
\bibinfo{author}{\bibfnamefont{M.}~\bibnamefont{Ajello}},
  \bibinfo{author}{\bibfnamefont{R.}~\bibnamefont{Romani}},
  \bibinfo{author}{\bibfnamefont{D.}~\bibnamefont{Gasparrini}},
  \bibinfo{author}{\bibfnamefont{M.}~\bibnamefont{Shaw}},
  \bibinfo{author}{\bibfnamefont{J.}~\bibnamefont{Bolmer}},
  \bibnamefont{et~al.}, \bibinfo{journal}{\apj} \textbf{\bibinfo{volume}{780}},
  \bibinfo{pages}{73} (\bibinfo{year}{2014}{\natexlab{a}}), \eprint{1310.0006}.

\bibitem[{\citenamefont{Ajello et~al.}(2014{\natexlab{b}})\citenamefont{Ajello,
  Gasparrini, S{\'a}nchez-Conde, Zaharijas et~al.}}]{Ajellog2014}
\bibinfo{author}{\bibfnamefont{M.}~\bibnamefont{Ajello}},
  \bibinfo{author}{\bibfnamefont{D.}~\bibnamefont{Gasparrini}},
  \bibinfo{author}{\bibfnamefont{M.}~\bibnamefont{S{\'a}nchez-Conde}},
  \bibinfo{author}{\bibfnamefont{G.}~\bibnamefont{Zaharijas}},
  \bibnamefont{et~al.} (\bibinfo{collaboration}{The Fermi-LAT}),
  \bibinfo{journal}{In preparation}  (\bibinfo{year}{2014}{\natexlab{b}}).

\bibitem[{\citenamefont{Di~Mauro
  et~al.}(2014{\natexlab{a}})\citenamefont{Di~Mauro, Donato, Lamanna, Sanchez,
  and Serpico}}]{DiMauro:2013zfa}
\bibinfo{author}{\bibfnamefont{M.}~\bibnamefont{Di~Mauro}},
  \bibinfo{author}{\bibfnamefont{F.}~\bibnamefont{Donato}},
  \bibinfo{author}{\bibfnamefont{G.}~\bibnamefont{Lamanna}},
  \bibinfo{author}{\bibfnamefont{D.}~\bibnamefont{Sanchez}}, \bibnamefont{and}
  \bibinfo{author}{\bibfnamefont{P.}~\bibnamefont{Serpico}},
  \bibinfo{journal}{\apj} \textbf{\bibinfo{volume}{786}}, \bibinfo{pages}{129}
  (\bibinfo{year}{2014}{\natexlab{a}}), \eprint{1311.5708}.

\bibitem[{\citenamefont{Di~Mauro
  et~al.}(2014{\natexlab{b}})\citenamefont{Di~Mauro, Calore, Donato, Ajello,
  and Latronico}}]{DiMauro:2013xta}
\bibinfo{author}{\bibfnamefont{M.}~\bibnamefont{Di~Mauro}},
  \bibinfo{author}{\bibfnamefont{F.}~\bibnamefont{Calore}},
  \bibinfo{author}{\bibfnamefont{F.}~\bibnamefont{Donato}},
  \bibinfo{author}{\bibfnamefont{M.}~\bibnamefont{Ajello}}, \bibnamefont{and}
  \bibinfo{author}{\bibfnamefont{L.}~\bibnamefont{Latronico}},
  \bibinfo{journal}{\apj} \textbf{\bibinfo{volume}{780}}, \bibinfo{pages}{161}
  (\bibinfo{year}{2014}{\natexlab{b}}), \eprint{1304.0908}.

\bibitem[{\citenamefont{{Inoue}}(2011)}]{2011ApJ...733...66I}
\bibinfo{author}{\bibfnamefont{Y.}~\bibnamefont{{Inoue}}},
  \bibinfo{journal}{\apj} \textbf{\bibinfo{volume}{733}}, \bibinfo{eid}{66}
  (\bibinfo{year}{2011}), \eprint{1103.3946}.

\bibitem[{\citenamefont{{Ackermann} et~al.}(2012)\citenamefont{{Ackermann},
  {Ajello}, {Allafort} et~al.}}]{2012ApJ...755..164A}
\bibinfo{author}{\bibfnamefont{M.}~\bibnamefont{{Ackermann}}},
  \bibinfo{author}{\bibfnamefont{M.}~\bibnamefont{{Ajello}}},
  \bibinfo{author}{\bibfnamefont{A.}~\bibnamefont{{Allafort}}},
  \bibnamefont{et~al.}, \bibinfo{journal}{\apj} \textbf{\bibinfo{volume}{755}},
  \bibinfo{pages}{164} (\bibinfo{year}{2012}), \eprint{1206.1346}.

\bibitem[{\citenamefont{{Fields} et~al.}(2010)\citenamefont{{Fields},
  {Pavlidou}, and {Prodanovi{\'c}}}}]{2010ApJ...722L.199F}
\bibinfo{author}{\bibfnamefont{B.~D.} \bibnamefont{{Fields}}},
  \bibinfo{author}{\bibfnamefont{V.}~\bibnamefont{{Pavlidou}}},
  \bibnamefont{and}
  \bibinfo{author}{\bibfnamefont{T.}~\bibnamefont{{Prodanovi{\'c}}}},
  \bibinfo{journal}{\apjl} \textbf{\bibinfo{volume}{722}},
  \bibinfo{pages}{L199} (\bibinfo{year}{2010}), \eprint{1003.3647}.

\bibitem[{\citenamefont{Cholis et~al.}(2014)\citenamefont{Cholis, Hooper, and
  McDermott}}]{Cholis:2013ena}
\bibinfo{author}{\bibfnamefont{I.}~\bibnamefont{Cholis}},
  \bibinfo{author}{\bibfnamefont{D.}~\bibnamefont{Hooper}}, \bibnamefont{and}
  \bibinfo{author}{\bibfnamefont{S.~D.} \bibnamefont{McDermott}},
  \bibinfo{journal}{JCAP} \textbf{\bibinfo{volume}{1402}}, \bibinfo{pages}{014}
  (\bibinfo{year}{2014}), \eprint{1312.0608}.

\bibitem[{\citenamefont{Calore et~al.}(2014{\natexlab{a}})\citenamefont{Calore,
  Di~Mauro, Donato, and Donato}}]{Calore:2014oga}
\bibinfo{author}{\bibfnamefont{F.}~\bibnamefont{Calore}},
  \bibinfo{author}{\bibfnamefont{M.}~\bibnamefont{Di~Mauro}},
  \bibinfo{author}{\bibfnamefont{F.}~\bibnamefont{Donato}}, \bibnamefont{and}
  \bibinfo{author}{\bibfnamefont{F.}~\bibnamefont{Donato}},
  \bibinfo{journal}{\apj} \textbf{\bibinfo{volume}{796}}, \bibinfo{pages}{1}
  (\bibinfo{year}{2014}{\natexlab{a}}), \eprint{1406.2706}.

\bibitem[{\citenamefont{{Abdo} et~al.}(2013)\citenamefont{{Abdo}, {Ajello},
  {Allafort}, {Baldini}, {Ballet}, {Barbiellini}, {Baring}, {Bastieri},
  {Belfiore}, {Bellazzini} et~al.}}]{2013ApJS..208...17A}
\bibinfo{author}{\bibfnamefont{A.~A.} \bibnamefont{{Abdo}}},
  \bibinfo{author}{\bibfnamefont{M.}~\bibnamefont{{Ajello}}},
  \bibinfo{author}{\bibfnamefont{A.}~\bibnamefont{{Allafort}}},
  \bibinfo{author}{\bibfnamefont{L.}~\bibnamefont{{Baldini}}},
  \bibinfo{author}{\bibfnamefont{J.}~\bibnamefont{{Ballet}}},
  \bibinfo{author}{\bibfnamefont{G.}~\bibnamefont{{Barbiellini}}},
  \bibinfo{author}{\bibfnamefont{M.~G.} \bibnamefont{{Baring}}},
  \bibinfo{author}{\bibfnamefont{D.}~\bibnamefont{{Bastieri}}},
  \bibinfo{author}{\bibfnamefont{A.}~\bibnamefont{{Belfiore}}},
  \bibinfo{author}{\bibfnamefont{R.}~\bibnamefont{{Bellazzini}}},
  \bibnamefont{et~al.}, \bibinfo{journal}{\apjs}
  \textbf{\bibinfo{volume}{208}}, \bibinfo{eid}{17} (\bibinfo{year}{2013}),
  \eprint{1305.4385}.

\bibitem[{\citenamefont{Hooper et~al.}(2013)\citenamefont{Hooper, Cholis,
  Linden, Siegal-Gaskins, and Slatyer}}]{Hooper:2013nhl}
\bibinfo{author}{\bibfnamefont{D.}~\bibnamefont{Hooper}},
  \bibinfo{author}{\bibfnamefont{I.}~\bibnamefont{Cholis}},
  \bibinfo{author}{\bibfnamefont{T.}~\bibnamefont{Linden}},
  \bibinfo{author}{\bibfnamefont{J.}~\bibnamefont{Siegal-Gaskins}},
  \bibnamefont{and} \bibinfo{author}{\bibfnamefont{T.}~\bibnamefont{Slatyer}},
  \bibinfo{journal}{\prd} \textbf{\bibinfo{volume}{D88}},
  \bibinfo{pages}{083009} (\bibinfo{year}{2013}), \eprint{1305.0830}.

\bibitem[{\citenamefont{{Gr{\'e}goire} and
  {Kn{\"o}dlseder}}(2013)}]{2013A&A...554A..62G}
\bibinfo{author}{\bibfnamefont{T.}~\bibnamefont{{Gr{\'e}goire}}}
  \bibnamefont{and}
  \bibinfo{author}{\bibfnamefont{J.}~\bibnamefont{{Kn{\"o}dlseder}}},
  \bibinfo{journal}{\aap} \textbf{\bibinfo{volume}{554}}, \bibinfo{eid}{A62}
  (\bibinfo{year}{2013}), \eprint{1305.1584}.

\bibitem[{\citenamefont{Di~Mauro
  et~al.}(2014{\natexlab{c}})\citenamefont{Di~Mauro, Cuoco, Donato, and
  Siegal-Gaskins}}]{DiMauro:2014wha}
\bibinfo{author}{\bibfnamefont{M.}~\bibnamefont{Di~Mauro}},
  \bibinfo{author}{\bibfnamefont{A.}~\bibnamefont{Cuoco}},
  \bibinfo{author}{\bibfnamefont{F.}~\bibnamefont{Donato}}, \bibnamefont{and}
  \bibinfo{author}{\bibfnamefont{J.~M.} \bibnamefont{Siegal-Gaskins}},
  \bibinfo{journal}{JCAP} \textbf{\bibinfo{volume}{1411}}, \bibinfo{pages}{021}
  (\bibinfo{year}{2014}{\natexlab{c}}), \eprint{1407.3275}.

\bibitem[{\citenamefont{Ackermann
  et~al.}(2012{\natexlab{a}})}]{2012PhRvD..85h3007A}
\bibinfo{author}{\bibfnamefont{M.}~\bibnamefont{Ackermann}}
  \bibnamefont{et~al.}, \bibinfo{journal}{\prd} \textbf{\bibinfo{volume}{85}},
  \bibinfo{pages}{083007} (\bibinfo{year}{2012}{\natexlab{a}}),
  \eprint{1202.2856}.

\bibitem[{\citenamefont{{Bengtsson} et~al.}(1990)\citenamefont{{Bengtsson},
  {Salati}, and {Silk}}}]{1990NuPhB.346..129B}
\bibinfo{author}{\bibfnamefont{H.}~\bibnamefont{{Bengtsson}}},
  \bibinfo{author}{\bibfnamefont{P.}~\bibnamefont{{Salati}}}, \bibnamefont{and}
  \bibinfo{author}{\bibfnamefont{J.}~\bibnamefont{{Silk}}},
  \bibinfo{journal}{Nuclear Phys. B} \textbf{\bibinfo{volume}{346}},
  \bibinfo{pages}{129} (\bibinfo{year}{1990}).

\bibitem[{\citenamefont{Bergstrom et~al.}(2001)\citenamefont{Bergstrom, Edsjo,
  and Ullio}}]{Bergstrom:2001jj}
\bibinfo{author}{\bibfnamefont{L.}~\bibnamefont{Bergstrom}},
  \bibinfo{author}{\bibfnamefont{J.}~\bibnamefont{Edsjo}}, \bibnamefont{and}
  \bibinfo{author}{\bibfnamefont{P.}~\bibnamefont{Ullio}},
  \bibinfo{journal}{Phys.Rev.Lett.} \textbf{\bibinfo{volume}{87}},
  \bibinfo{pages}{251301} (\bibinfo{year}{2001}), \eprint{astro-ph/0105048}.

\bibitem[{\citenamefont{Ullio et~al.}(2002)\citenamefont{Ullio, Bergstrom,
  Edsjo, and Lacey}}]{Ullio:2002pj}
\bibinfo{author}{\bibfnamefont{P.}~\bibnamefont{Ullio}},
  \bibinfo{author}{\bibfnamefont{L.}~\bibnamefont{Bergstrom}},
  \bibinfo{author}{\bibfnamefont{J.}~\bibnamefont{Edsjo}}, \bibnamefont{and}
  \bibinfo{author}{\bibfnamefont{C.~G.} \bibnamefont{Lacey}},
  \bibinfo{journal}{\prd} \textbf{\bibinfo{volume}{66}},
  \bibinfo{pages}{123502} (\bibinfo{year}{2002}), \eprint{astro-ph/0207125}.

\bibitem[{\citenamefont{Bottino et~al.}(2004)\citenamefont{Bottino, Donato,
  Fornengo, and Scopel}}]{Bottino:2004qi}
\bibinfo{author}{\bibfnamefont{A.}~\bibnamefont{Bottino}},
  \bibinfo{author}{\bibfnamefont{F.}~\bibnamefont{Donato}},
  \bibinfo{author}{\bibfnamefont{N.}~\bibnamefont{Fornengo}}, \bibnamefont{and}
  \bibinfo{author}{\bibfnamefont{S.}~\bibnamefont{Scopel}},
  \bibinfo{journal}{\prd} \textbf{\bibinfo{volume}{70}},
  \bibinfo{pages}{015005} (\bibinfo{year}{2004}), \eprint{astro-ph/0401186}.

\bibitem[{\citenamefont{Bringmann and Weniger}(2012)}]{Bringmann:2012ez}
\bibinfo{author}{\bibfnamefont{T.}~\bibnamefont{Bringmann}} \bibnamefont{and}
  \bibinfo{author}{\bibfnamefont{C.}~\bibnamefont{Weniger}},
  \bibinfo{journal}{Phys.Dark Univ.} \textbf{\bibinfo{volume}{1}},
  \bibinfo{pages}{194} (\bibinfo{year}{2012}), \eprint{1208.5481}.

\bibitem[{\citenamefont{{Cirelli} et~al.}(2010)\citenamefont{{Cirelli},
  {Panci}, and {Serpico}}}]{2010NuPhB.840..284C}
\bibinfo{author}{\bibfnamefont{M.}~\bibnamefont{{Cirelli}}},
  \bibinfo{author}{\bibfnamefont{P.}~\bibnamefont{{Panci}}}, \bibnamefont{and}
  \bibinfo{author}{\bibfnamefont{P.~D.} \bibnamefont{{Serpico}}},
  \bibinfo{journal}{Nuclear Phys. B} \textbf{\bibinfo{volume}{840}},
  \bibinfo{pages}{284} (\bibinfo{year}{2010}), \eprint{0912.0663}.

\bibitem[{\citenamefont{{Calore} et~al.}(2012)\citenamefont{{Calore}, {de
  Romeri}, and {Donato}}}]{2012PhRvD..85b3004C}
\bibinfo{author}{\bibfnamefont{F.}~\bibnamefont{{Calore}}},
  \bibinfo{author}{\bibfnamefont{V.}~\bibnamefont{{de Romeri}}},
  \bibnamefont{and} \bibinfo{author}{\bibfnamefont{F.}~\bibnamefont{{Donato}}},
  \bibinfo{journal}{\prd} \textbf{\bibinfo{volume}{85}},
  \bibinfo{pages}{023004} (\bibinfo{year}{2012}), \eprint{1105.4230}.

\bibitem[{\citenamefont{Abazajian et~al.}(2012)\citenamefont{Abazajian,
  Blanchet, and Harding}}]{Abazajian:2010zb}
\bibinfo{author}{\bibfnamefont{K.~N.} \bibnamefont{Abazajian}},
  \bibinfo{author}{\bibfnamefont{S.}~\bibnamefont{Blanchet}}, \bibnamefont{and}
  \bibinfo{author}{\bibfnamefont{J.~P.} \bibnamefont{Harding}},
  \bibinfo{journal}{\prd} \textbf{\bibinfo{volume}{85}},
  \bibinfo{pages}{043509} (\bibinfo{year}{2012}), \eprint{1011.5090}.

\bibitem[{\citenamefont{Bringmann et~al.}(2014)\citenamefont{Bringmann, Calore,
  Di~Mauro, and Donato}}]{Calore:2013yia}
\bibinfo{author}{\bibfnamefont{T.}~\bibnamefont{Bringmann}},
  \bibinfo{author}{\bibfnamefont{F.}~\bibnamefont{Calore}},
  \bibinfo{author}{\bibfnamefont{M.}~\bibnamefont{Di~Mauro}}, \bibnamefont{and}
  \bibinfo{author}{\bibfnamefont{F.}~\bibnamefont{Donato}},
  \bibinfo{journal}{\prd} \textbf{\bibinfo{volume}{89}},
  \bibinfo{pages}{023012} (\bibinfo{year}{2014}), \eprint{1303.3284}.

\bibitem[{\citenamefont{Ackermann
  et~al.}(2012{\natexlab{b}})}]{Ackermann:2012rg}
\bibinfo{author}{\bibfnamefont{M.}~\bibnamefont{Ackermann}}
  \bibnamefont{et~al.}, \bibinfo{journal}{\apj} \textbf{\bibinfo{volume}{761}},
  \bibinfo{pages}{91} (\bibinfo{year}{2012}{\natexlab{b}}), \eprint{1205.6474}.

\bibitem[{\citenamefont{Ackermann
  et~al.}(2014{\natexlab{b}})\citenamefont{Ackermann, Ajello, Albert, Baldini,
  Barbiellini et~al.}}]{Ackermanna2014}
\bibinfo{author}{\bibfnamefont{M.}~\bibnamefont{Ackermann}},
  \bibinfo{author}{\bibfnamefont{M.}~\bibnamefont{Ajello}},
  \bibinfo{author}{\bibfnamefont{A.}~\bibnamefont{Albert}},
  \bibinfo{author}{\bibfnamefont{L.}~\bibnamefont{Baldini}},
  \bibinfo{author}{\bibfnamefont{G.}~\bibnamefont{Barbiellini}},
  \bibnamefont{et~al.} (\bibinfo{collaboration}{The Fermi-LAT}),
  \bibinfo{journal}{In preparation}  (\bibinfo{year}{2014}{\natexlab{b}}).

\bibitem[{\citenamefont{Massari et~al.}(2014)\citenamefont{Massari, Izaguirre,
  Essig, Albert, and Bloom}}]{Massarii2014}
\bibinfo{author}{\bibfnamefont{A.}~\bibnamefont{Massari}},
  \bibinfo{author}{\bibfnamefont{E.}~\bibnamefont{Izaguirre}},
  \bibinfo{author}{\bibfnamefont{R.}~\bibnamefont{Essig}},
  \bibinfo{author}{\bibfnamefont{A.}~\bibnamefont{Albert}}, \bibnamefont{and}
  \bibinfo{author}{\bibfnamefont{E.}~\bibnamefont{Bloom}}
  (\bibinfo{collaboration}{The Fermi-LAT}), \bibinfo{journal}{In preparation}
  (\bibinfo{year}{2014}).

\bibitem[{\citenamefont{{Angel} and {Stockman}}(1980)}]{1980ARA&A..18..321A}
\bibinfo{author}{\bibfnamefont{J.~R.~P.} \bibnamefont{{Angel}}}
  \bibnamefont{and} \bibinfo{author}{\bibfnamefont{H.~S.}
  \bibnamefont{{Stockman}}}, \bibinfo{journal}{\araa}
  \textbf{\bibinfo{volume}{18}}, \bibinfo{pages}{321} (\bibinfo{year}{1980}).

\bibitem[{\citenamefont{{Nolan} et~al.}(2012)}]{2012yCat..21990031N}
\bibinfo{author}{\bibfnamefont{P.~L.} \bibnamefont{{Nolan}}}
  \bibnamefont{et~al.}, \bibinfo{journal}{VizieR Online Data Catalog}
  \textbf{\bibinfo{volume}{219}}, \bibinfo{pages}{90031}
  (\bibinfo{year}{2012}).

\bibitem[{\citenamefont{{Ackermann} et~al.}(2011)}]{secondcatalogAGN}
\bibinfo{author}{\bibfnamefont{M.}~\bibnamefont{{Ackermann}}}
  \bibnamefont{et~al.}, \bibinfo{journal}{\apj} \textbf{\bibinfo{volume}{743}},
  \bibinfo{eid}{171} (\bibinfo{year}{2011}).

\bibitem[{\citenamefont{{Abdo}
  et~al.}(2010{\natexlab{b}})}]{2010ApJ...720..435A}
\bibinfo{author}{\bibfnamefont{A.~A.} \bibnamefont{{Abdo}}}
  \bibnamefont{et~al.}, \bibinfo{journal}{\apj} \textbf{\bibinfo{volume}{720}},
  \bibinfo{pages}{435} (\bibinfo{year}{2010}{\natexlab{b}}),
  \eprint{1003.0895}.

\bibitem[{\citenamefont{{Ajello} et~al.}(2012)\citenamefont{{Ajello}, {Shaw},
  {Romani}, {Dermer}, {Costamante}, {King}, {Max-Moerbeck}, {Readhead},
  {Reimer}, {Richards} et~al.}}]{2012ApJ...751..108A}
\bibinfo{author}{\bibfnamefont{M.}~\bibnamefont{{Ajello}}},
  \bibinfo{author}{\bibfnamefont{M.~S.} \bibnamefont{{Shaw}}},
  \bibinfo{author}{\bibfnamefont{R.~W.} \bibnamefont{{Romani}}},
  \bibinfo{author}{\bibfnamefont{C.~D.} \bibnamefont{{Dermer}}},
  \bibinfo{author}{\bibfnamefont{L.}~\bibnamefont{{Costamante}}},
  \bibinfo{author}{\bibfnamefont{O.~G.} \bibnamefont{{King}}},
  \bibinfo{author}{\bibfnamefont{W.}~\bibnamefont{{Max-Moerbeck}}},
  \bibinfo{author}{\bibfnamefont{A.}~\bibnamefont{{Readhead}}},
  \bibinfo{author}{\bibfnamefont{A.}~\bibnamefont{{Reimer}}},
  \bibinfo{author}{\bibfnamefont{J.~L.} \bibnamefont{{Richards}}},
  \bibnamefont{et~al.}, \bibinfo{journal}{\apj} \textbf{\bibinfo{volume}{751}},
  \bibinfo{pages}{108} (\bibinfo{year}{2012}), \eprint{1110.3787}.

\bibitem[{\citenamefont{{Abdo} et~al.}(2010{\natexlab{c}})\citenamefont{{Abdo},
  {Ackermann}, {Ajello}, {Allafort}, {Antolini}, {Atwood}, {Axelsson},
  {Baldini}, {Ballet}, {Barbiellini} et~al.}}]{1FGLC}
\bibinfo{author}{\bibfnamefont{A.~A.} \bibnamefont{{Abdo}}},
  \bibinfo{author}{\bibfnamefont{M.}~\bibnamefont{{Ackermann}}},
  \bibinfo{author}{\bibfnamefont{M.}~\bibnamefont{{Ajello}}},
  \bibinfo{author}{\bibfnamefont{A.}~\bibnamefont{{Allafort}}},
  \bibinfo{author}{\bibfnamefont{E.}~\bibnamefont{{Antolini}}},
  \bibinfo{author}{\bibfnamefont{W.~B.} \bibnamefont{{Atwood}}},
  \bibinfo{author}{\bibfnamefont{M.}~\bibnamefont{{Axelsson}}},
  \bibinfo{author}{\bibfnamefont{L.}~\bibnamefont{{Baldini}}},
  \bibinfo{author}{\bibfnamefont{J.}~\bibnamefont{{Ballet}}},
  \bibinfo{author}{\bibfnamefont{G.}~\bibnamefont{{Barbiellini}}},
  \bibnamefont{et~al.}, \bibinfo{journal}{Astrophys. J. S.S.}
  \textbf{\bibinfo{volume}{188}}, \bibinfo{pages}{405}
  (\bibinfo{year}{2010}{\natexlab{c}}), \eprint{1002.2280}.

\bibitem[{\citenamefont{{Padovani} and {Giommi}}(1995)}]{1995ApJ...444..567P}
\bibinfo{author}{\bibfnamefont{P.}~\bibnamefont{{Padovani}}} \bibnamefont{and}
  \bibinfo{author}{\bibfnamefont{P.}~\bibnamefont{{Giommi}}},
  \bibinfo{journal}{\apj} \textbf{\bibinfo{volume}{444}}, \bibinfo{pages}{567}
  (\bibinfo{year}{1995}), \eprint{astro-ph/9412073}.

\bibitem[{\citenamefont{{Abdo} et~al.}(2010{\natexlab{d}})\citenamefont{{Abdo},
  {Ackermann}, {Agudo}, {Ajello}, {Aller}, {Aller}, {Angelakis}, {Arkharov},
  {Axelsson}, {Bach} et~al.}}]{2010ApJ...716...30A}
\bibinfo{author}{\bibfnamefont{A.~A.} \bibnamefont{{Abdo}}},
  \bibinfo{author}{\bibfnamefont{M.}~\bibnamefont{{Ackermann}}},
  \bibinfo{author}{\bibfnamefont{I.}~\bibnamefont{{Agudo}}},
  \bibinfo{author}{\bibfnamefont{M.}~\bibnamefont{{Ajello}}},
  \bibinfo{author}{\bibfnamefont{H.~D.} \bibnamefont{{Aller}}},
  \bibinfo{author}{\bibfnamefont{M.~F.} \bibnamefont{{Aller}}},
  \bibinfo{author}{\bibfnamefont{E.}~\bibnamefont{{Angelakis}}},
  \bibinfo{author}{\bibfnamefont{A.~A.} \bibnamefont{{Arkharov}}},
  \bibinfo{author}{\bibfnamefont{M.}~\bibnamefont{{Axelsson}}},
  \bibinfo{author}{\bibfnamefont{U.}~\bibnamefont{{Bach}}},
  \bibnamefont{et~al.}, \bibinfo{journal}{\apj} \textbf{\bibinfo{volume}{716}},
  \bibinfo{pages}{30} (\bibinfo{year}{2010}{\natexlab{d}}), \eprint{0912.2040}.

\bibitem[{\citenamefont{Abdo et~al.}(2010)}]{Abdo:2010ru}
\bibinfo{author}{\bibfnamefont{A.}~\bibnamefont{Abdo}} \bibnamefont{et~al.},
  \bibinfo{journal}{Astrophys.J.Suppl.} \textbf{\bibinfo{volume}{188}},
  \bibinfo{pages}{405} (\bibinfo{year}{2010}), \eprint{1002.2280}.

\bibitem[{\citenamefont{Abdo et~al.}(2013)}]{TheFermi-LAT:2013xza}
\bibinfo{author}{\bibfnamefont{A.}~\bibnamefont{Abdo}} \bibnamefont{et~al.}
  (\bibinfo{year}{2013}), \eprint{1306.6772}.

\bibitem[{\citenamefont{{Wakely} and {Horan}}(2013)}]{tevcat}
\bibinfo{author}{\bibfnamefont{S.}~\bibnamefont{{Wakely}}} \bibnamefont{and}
  \bibinfo{author}{\bibfnamefont{D.}~\bibnamefont{{Horan}}},
  \bibinfo{journal}{\url{http://tevcat.uchicago.edu}}  (\bibinfo{year}{2013}).

\bibitem[{\citenamefont{{Abdo}
  et~al.}(2010{\natexlab{a}})}]{2010ApJ...709L.152A}
\bibinfo{author}{\bibfnamefont{A.~A.} \bibnamefont{{Abdo}}}
  \bibnamefont{et~al.}, \bibinfo{journal}{\apjl}
  \textbf{\bibinfo{volume}{709}}, \bibinfo{pages}{L152}
  (\bibinfo{year}{2010}{\natexlab{a}}), \eprint{0911.5327}.

\bibitem[{\citenamefont{{Lacki} et~al.}(2011)\citenamefont{{Lacki}, {Thompson},
  {Quataert}, {Loeb}, and {Waxman}}}]{2011ApJ...734..107L}
\bibinfo{author}{\bibfnamefont{B.~C.} \bibnamefont{{Lacki}}},
  \bibinfo{author}{\bibfnamefont{T.~A.} \bibnamefont{{Thompson}}},
  \bibinfo{author}{\bibfnamefont{E.}~\bibnamefont{{Quataert}}},
  \bibinfo{author}{\bibfnamefont{A.}~\bibnamefont{{Loeb}}}, \bibnamefont{and}
  \bibinfo{author}{\bibfnamefont{E.}~\bibnamefont{{Waxman}}},
  \bibinfo{journal}{\apj} \textbf{\bibinfo{volume}{734}}, \bibinfo{eid}{107}
  (\bibinfo{year}{2011}), \eprint{1003.3257}.

\bibitem[{\citenamefont{{Chakraborty} and
  {Fields}}(2013)}]{2013ApJ...773..104C}
\bibinfo{author}{\bibfnamefont{N.}~\bibnamefont{{Chakraborty}}}
  \bibnamefont{and} \bibinfo{author}{\bibfnamefont{B.~D.}
  \bibnamefont{{Fields}}}, \bibinfo{journal}{\apj}
  \textbf{\bibinfo{volume}{773}}, \bibinfo{eid}{104} (\bibinfo{year}{2013}),
  \eprint{1206.0770}.

\bibitem[{\citenamefont{{Tamborra} et~al.}(2014)\citenamefont{{Tamborra},
  {Ando}, and {Murase}}}]{2014JCAP...09..043T}
\bibinfo{author}{\bibfnamefont{I.}~\bibnamefont{{Tamborra}}},
  \bibinfo{author}{\bibfnamefont{S.}~\bibnamefont{{Ando}}}, \bibnamefont{and}
  \bibinfo{author}{\bibfnamefont{K.}~\bibnamefont{{Murase}}},
  \bibinfo{journal}{\jcap} \textbf{\bibinfo{volume}{9}}, \bibinfo{eid}{043}
  (\bibinfo{year}{2014}), \eprint{1404.1189}.

\bibitem[{\citenamefont{{Berezinsky} et~al.}(2011)\citenamefont{{Berezinsky},
  {Gazizov}, {Kachelrie{\ss}}, and {Ostapchenko}}}]{2011PhLB..695...13B}
\bibinfo{author}{\bibfnamefont{V.}~\bibnamefont{{Berezinsky}}},
  \bibinfo{author}{\bibfnamefont{A.}~\bibnamefont{{Gazizov}}},
  \bibinfo{author}{\bibfnamefont{M.}~\bibnamefont{{Kachelrie{\ss}}}},
  \bibnamefont{and}
  \bibinfo{author}{\bibfnamefont{S.}~\bibnamefont{{Ostapchenko}}},
  \bibinfo{journal}{Physics Letters B} \textbf{\bibinfo{volume}{695}},
  \bibinfo{pages}{13} (\bibinfo{year}{2011}), \eprint{1003.1496}.

\bibitem[{\citenamefont{{Gabici} and {Blasi}}(2004)}]{2004APh....20..579G}
\bibinfo{author}{\bibfnamefont{S.}~\bibnamefont{{Gabici}}} \bibnamefont{and}
  \bibinfo{author}{\bibfnamefont{P.}~\bibnamefont{{Blasi}}},
  \bibinfo{journal}{Astropart.Phys.} \textbf{\bibinfo{volume}{20}},
  \bibinfo{pages}{579} (\bibinfo{year}{2004}), \eprint{astro-ph/0306369}.

\bibitem[{\citenamefont{{Zandanel} et~al.}(2014)\citenamefont{{Zandanel},
  {Tamborra}, {Gabici}, and {Ando}}}]{2014arXiv1410.8697Z}
\bibinfo{author}{\bibfnamefont{F.}~\bibnamefont{{Zandanel}}},
  \bibinfo{author}{\bibfnamefont{I.}~\bibnamefont{{Tamborra}}},
  \bibinfo{author}{\bibfnamefont{S.}~\bibnamefont{{Gabici}}}, \bibnamefont{and}
  \bibinfo{author}{\bibfnamefont{S.}~\bibnamefont{{Ando}}},
  \bibinfo{journal}{ArXiv e-prints}  (\bibinfo{year}{2014}),
  \eprint{1410.8697}.

\bibitem[{\citenamefont{{Abdo}
  et~al.}(2010{\natexlab{b}})}]{2010ApJBlazarFermi}
\bibinfo{author}{\bibfnamefont{A.~A.} \bibnamefont{{Abdo}}}
  \bibnamefont{et~al.}, \bibinfo{journal}{\apj} \textbf{\bibinfo{volume}{720}},
  \bibinfo{pages}{435} (\bibinfo{year}{2010}{\natexlab{b}}),
  \eprint{1003.0895}.

\bibitem[{\citenamefont{{Finke} et~al.}(2010)\citenamefont{{Finke}, {Razzaque},
  and {Dermer}}}]{2010ApJ...712..238F}
\bibinfo{author}{\bibfnamefont{J.~D.} \bibnamefont{{Finke}}},
  \bibinfo{author}{\bibfnamefont{S.}~\bibnamefont{{Razzaque}}},
  \bibnamefont{and} \bibinfo{author}{\bibfnamefont{C.~D.}
  \bibnamefont{{Dermer}}}, \bibinfo{journal}{\apj}
  \textbf{\bibinfo{volume}{712}}, \bibinfo{pages}{238} (\bibinfo{year}{2010}),
  \eprint{0905.1115}.

\bibitem[{\citenamefont{Ackermann
  et~al.}(2012{\natexlab{c}})}]{Ackermann:2012sza}
\bibinfo{author}{\bibfnamefont{M.}~\bibnamefont{Ackermann}}
  \bibnamefont{et~al.} (\bibinfo{collaboration}{The Fermi-LAT}),
  \bibinfo{journal}{Science} \textbf{\bibinfo{volume}{338}},
  \bibinfo{pages}{1190} (\bibinfo{year}{2012}{\natexlab{c}}),
  \eprint{1211.1671}.

\bibitem[{\citenamefont{Abramowski et~al.}(2012)}]{Abramowski:2012ry}
\bibinfo{author}{\bibfnamefont{A.}~\bibnamefont{Abramowski}}
  \bibnamefont{et~al.} (\bibinfo{collaboration}{The H.E.S.S.}),
  \bibinfo{journal}{\aap} \textbf{\bibinfo{volume}{550}}
  (\bibinfo{year}{2012}), \eprint{1212.3409}.

\bibitem[{\citenamefont{Fornengo et~al.}(2015)\citenamefont{Fornengo, Perotto,
  Regis, and Camera}}]{Fornengo:2014cya}
\bibinfo{author}{\bibfnamefont{N.}~\bibnamefont{Fornengo}},
  \bibinfo{author}{\bibfnamefont{L.}~\bibnamefont{Perotto}},
  \bibinfo{author}{\bibfnamefont{M.}~\bibnamefont{Regis}}, \bibnamefont{and}
  \bibinfo{author}{\bibfnamefont{S.}~\bibnamefont{Camera}},
  \bibinfo{journal}{Astrophys.J.} \textbf{\bibinfo{volume}{802}},
  \bibinfo{pages}{L1} (\bibinfo{year}{2015}), \eprint{1410.4997}.

\bibitem[{\citenamefont{Taylor and Silk}(2003)}]{Taylor:2002zd}
\bibinfo{author}{\bibfnamefont{J.~E.} \bibnamefont{Taylor}} \bibnamefont{and}
  \bibinfo{author}{\bibfnamefont{J.}~\bibnamefont{Silk}},
  \bibinfo{journal}{\mnras} \textbf{\bibinfo{volume}{339}},
  \bibinfo{pages}{505} (\bibinfo{year}{2003}), \eprint{astro-ph/0207299}.

\bibitem[{\citenamefont{Papucci and Strumia}(2010)}]{Papucci:2009gd}
\bibinfo{author}{\bibfnamefont{M.}~\bibnamefont{Papucci}} \bibnamefont{and}
  \bibinfo{author}{\bibfnamefont{A.}~\bibnamefont{Strumia}},
  \bibinfo{journal}{JCAP} \textbf{\bibinfo{volume}{1003}}, \bibinfo{pages}{014}
  (\bibinfo{year}{2010}), \eprint{0912.0742}.

\bibitem[{\citenamefont{Cirelli et~al.}(2010)\citenamefont{Cirelli, Panci, and
  Serpico}}]{Cirelli:2009dv}
\bibinfo{author}{\bibfnamefont{M.}~\bibnamefont{Cirelli}},
  \bibinfo{author}{\bibfnamefont{P.}~\bibnamefont{Panci}}, \bibnamefont{and}
  \bibinfo{author}{\bibfnamefont{P.~D.} \bibnamefont{Serpico}},
  \bibinfo{journal}{Nucl.Phys.} \textbf{\bibinfo{volume}{B840}},
  \bibinfo{pages}{284} (\bibinfo{year}{2010}), \eprint{0912.0663}.

\bibitem[{\citenamefont{Baxter et~al.}(2010)\citenamefont{Baxter, Dodelson,
  Koushiappas, and Strigari}}]{Baxter:2010fr}
\bibinfo{author}{\bibfnamefont{E.~J.} \bibnamefont{Baxter}},
  \bibinfo{author}{\bibfnamefont{S.}~\bibnamefont{Dodelson}},
  \bibinfo{author}{\bibfnamefont{S.~M.} \bibnamefont{Koushiappas}},
  \bibnamefont{and} \bibinfo{author}{\bibfnamefont{L.~E.}
  \bibnamefont{Strigari}}, \bibinfo{journal}{\prd}
  \textbf{\bibinfo{volume}{82}}, \bibinfo{pages}{123511}
  (\bibinfo{year}{2010}), \eprint{1006.2399}.

\bibitem[{\citenamefont{Blanchet and Lavalle}(2012)}]{Blanchet:2012vq}
\bibinfo{author}{\bibfnamefont{S.}~\bibnamefont{Blanchet}} \bibnamefont{and}
  \bibinfo{author}{\bibfnamefont{J.}~\bibnamefont{Lavalle}},
  \bibinfo{journal}{JCAP} \textbf{\bibinfo{volume}{1211}}, \bibinfo{pages}{021}
  (\bibinfo{year}{2012}), \eprint{1207.2476}.

\bibitem[{\citenamefont{Calore et~al.}(2014{\natexlab{b}})\citenamefont{Calore,
  De~Romeri, Di~Mauro, Donato, Herpich et~al.}}]{Calore:2014hna}
\bibinfo{author}{\bibfnamefont{F.}~\bibnamefont{Calore}},
  \bibinfo{author}{\bibfnamefont{V.}~\bibnamefont{De~Romeri}},
  \bibinfo{author}{\bibfnamefont{M.}~\bibnamefont{Di~Mauro}},
  \bibinfo{author}{\bibfnamefont{F.}~\bibnamefont{Donato}},
  \bibinfo{author}{\bibfnamefont{J.}~\bibnamefont{Herpich}},
  \bibnamefont{et~al.}, \bibinfo{journal}{\mnras}
  \textbf{\bibinfo{volume}{442}}, \bibinfo{pages}{1151}
  (\bibinfo{year}{2014}{\natexlab{b}}), \eprint{1402.0512}.

\bibitem[{\citenamefont{Bergstrom et~al.}(1998)\citenamefont{Bergstrom, Ullio,
  and Buckley}}]{Bergstrom:1997fj}
\bibinfo{author}{\bibfnamefont{L.}~\bibnamefont{Bergstrom}},
  \bibinfo{author}{\bibfnamefont{P.}~\bibnamefont{Ullio}}, \bibnamefont{and}
  \bibinfo{author}{\bibfnamefont{J.~H.} \bibnamefont{Buckley}},
  \bibinfo{journal}{Astropart.Phys.} \textbf{\bibinfo{volume}{9}},
  \bibinfo{pages}{137} (\bibinfo{year}{1998}), \eprint{astro-ph/9712318}.

\bibitem[{\citenamefont{{Cirelli} and {Panci}}(2009)}]{2009NuPhB.821..399C}
\bibinfo{author}{\bibfnamefont{M.}~\bibnamefont{{Cirelli}}} \bibnamefont{and}
  \bibinfo{author}{\bibfnamefont{P.}~\bibnamefont{{Panci}}},
  \bibinfo{journal}{Nuclear Phys. B} \textbf{\bibinfo{volume}{821}},
  \bibinfo{pages}{399} (\bibinfo{year}{2009}), \eprint{0904.3830}.

\bibitem[{\citenamefont{Cirelli et~al.}(2011)\citenamefont{Cirelli, Corcella,
  Hektor, Hutsi, Kadastik et~al.}}]{Cirelli:2010xx}
\bibinfo{author}{\bibfnamefont{M.}~\bibnamefont{Cirelli}},
  \bibinfo{author}{\bibfnamefont{G.}~\bibnamefont{Corcella}},
  \bibinfo{author}{\bibfnamefont{A.}~\bibnamefont{Hektor}},
  \bibinfo{author}{\bibfnamefont{G.}~\bibnamefont{Hutsi}},
  \bibinfo{author}{\bibfnamefont{M.}~\bibnamefont{Kadastik}},
  \bibnamefont{et~al.}, \bibinfo{journal}{JCAP}
  \textbf{\bibinfo{volume}{1103}}, \bibinfo{pages}{051} (\bibinfo{year}{2011}),
  \eprint{1012.4515}.

\bibitem[{\citenamefont{{Blumenthal} and {Gould}}(1970)}]{1970RvMP...42..237B}
\bibinfo{author}{\bibfnamefont{G.~R.} \bibnamefont{{Blumenthal}}}
  \bibnamefont{and} \bibinfo{author}{\bibfnamefont{R.~J.}
  \bibnamefont{{Gould}}}, \bibinfo{journal}{RevModPhys.}
  \textbf{\bibinfo{volume}{42}}, \bibinfo{pages}{237} (\bibinfo{year}{1970}).

\bibitem[{\citenamefont{{Longair}}(1994)}]{1994hea2.book.....L}
\bibinfo{author}{\bibfnamefont{M.~S.} \bibnamefont{{Longair}}},
  \emph{\bibinfo{title}{{High energy astrophysics. Volume 2. Stars, the Galaxy
  and the interstellar medium.}}} (\bibinfo{year}{1994}).

\bibitem[{\citenamefont{Donato et~al.}(2004)\citenamefont{Donato, Fornengo,
  Maurin, and Salati}}]{Donato:2003xg}
\bibinfo{author}{\bibfnamefont{F.}~\bibnamefont{Donato}},
  \bibinfo{author}{\bibfnamefont{N.}~\bibnamefont{Fornengo}},
  \bibinfo{author}{\bibfnamefont{D.}~\bibnamefont{Maurin}}, \bibnamefont{and}
  \bibinfo{author}{\bibfnamefont{P.}~\bibnamefont{Salati}},
  \bibinfo{journal}{\prd} \textbf{\bibinfo{volume}{69}},
  \bibinfo{pages}{063501} (\bibinfo{year}{2004}), \eprint{astro-ph/0306207}.

\bibitem[{\citenamefont{Sjostrand et~al.}(2008)\citenamefont{Sjostrand, Mrenna,
  and Skands}}]{Sjostrand:2007gs}
\bibinfo{author}{\bibfnamefont{T.}~\bibnamefont{Sjostrand}},
  \bibinfo{author}{\bibfnamefont{S.}~\bibnamefont{Mrenna}}, \bibnamefont{and}
  \bibinfo{author}{\bibfnamefont{P.~Z.} \bibnamefont{Skands}},
  \bibinfo{journal}{Comput.Phys.Commun.} \textbf{\bibinfo{volume}{178}},
  \bibinfo{pages}{852} (\bibinfo{year}{2008}), \eprint{0710.3820}.

\bibitem[{\citenamefont{{Salucci} et~al.}(2010)\citenamefont{{Salucci},
  {Nesti}, {Gentile}, and {Frigerio Martins}}}]{2010A&A...523A..83S}
\bibinfo{author}{\bibfnamefont{P.}~\bibnamefont{{Salucci}}},
  \bibinfo{author}{\bibfnamefont{F.}~\bibnamefont{{Nesti}}},
  \bibinfo{author}{\bibfnamefont{G.}~\bibnamefont{{Gentile}}},
  \bibnamefont{and} \bibinfo{author}{\bibfnamefont{C.}~\bibnamefont{{Frigerio
  Martins}}}, \bibinfo{journal}{\aap} \textbf{\bibinfo{volume}{523}},
  \bibinfo{pages}{A83} (\bibinfo{year}{2010}), \eprint{1003.3101}.

\bibitem[{\citenamefont{{Catena} and {Ullio}}(2010)}]{2010JCAP...08..004C}
\bibinfo{author}{\bibfnamefont{R.}~\bibnamefont{{Catena}}} \bibnamefont{and}
  \bibinfo{author}{\bibfnamefont{P.}~\bibnamefont{{Ullio}}},
  \bibinfo{journal}{\jcap} \textbf{\bibinfo{volume}{8}}, \bibinfo{eid}{004}
  (\bibinfo{year}{2010}), \eprint{0907.0018}.

\bibitem[{\citenamefont{Gillessen et~al.}(2009)\citenamefont{Gillessen,
  Eisenhauer, Trippe, Alexander, Genzel et~al.}}]{Gillessen:2008qv}
\bibinfo{author}{\bibfnamefont{S.}~\bibnamefont{Gillessen}},
  \bibinfo{author}{\bibfnamefont{F.}~\bibnamefont{Eisenhauer}},
  \bibinfo{author}{\bibfnamefont{S.}~\bibnamefont{Trippe}},
  \bibinfo{author}{\bibfnamefont{T.}~\bibnamefont{Alexander}},
  \bibinfo{author}{\bibfnamefont{R.}~\bibnamefont{Genzel}},
  \bibnamefont{et~al.}, \bibinfo{journal}{\apj} \textbf{\bibinfo{volume}{692}},
  \bibinfo{pages}{1075} (\bibinfo{year}{2009}), \eprint{0810.4674}.

\bibitem[{\citenamefont{{Bovy} et~al.}(2009)\citenamefont{{Bovy}, {Hogg}, and
  {Rix}}}]{2009ApJ...704.1704B}
\bibinfo{author}{\bibfnamefont{J.}~\bibnamefont{{Bovy}}},
  \bibinfo{author}{\bibfnamefont{D.~W.} \bibnamefont{{Hogg}}},
  \bibnamefont{and} \bibinfo{author}{\bibfnamefont{H.-W.} \bibnamefont{{Rix}}},
  \bibinfo{journal}{\apj} \textbf{\bibinfo{volume}{704}}, \bibinfo{pages}{1704}
  (\bibinfo{year}{2009}), \eprint{0907.5423}.

\bibitem[{\citenamefont{Ghez et~al.}(2008)\citenamefont{Ghez, Salim, Weinberg,
  Lu, Do et~al.}}]{Ghez:2008ms}
\bibinfo{author}{\bibfnamefont{A.}~\bibnamefont{Ghez}},
  \bibinfo{author}{\bibfnamefont{S.}~\bibnamefont{Salim}},
  \bibinfo{author}{\bibfnamefont{N.}~\bibnamefont{Weinberg}},
  \bibinfo{author}{\bibfnamefont{J.}~\bibnamefont{Lu}},
  \bibinfo{author}{\bibfnamefont{T.}~\bibnamefont{Do}}, \bibnamefont{et~al.},
  \bibinfo{journal}{\apj} \textbf{\bibinfo{volume}{689}}, \bibinfo{pages}{1044}
  (\bibinfo{year}{2008}), \eprint{0808.2870}.

\bibitem[{\citenamefont{Bergstrom et~al.}(2013)\citenamefont{Bergstrom,
  Bringmann, Cholis, Hooper, and Weniger}}]{Bergstrom:2013jra}
\bibinfo{author}{\bibfnamefont{L.}~\bibnamefont{Bergstrom}},
  \bibinfo{author}{\bibfnamefont{T.}~\bibnamefont{Bringmann}},
  \bibinfo{author}{\bibfnamefont{I.}~\bibnamefont{Cholis}},
  \bibinfo{author}{\bibfnamefont{D.}~\bibnamefont{Hooper}}, \bibnamefont{and}
  \bibinfo{author}{\bibfnamefont{C.}~\bibnamefont{Weniger}},
  \bibinfo{journal}{Phys.Rev.Lett.} \textbf{\bibinfo{volume}{111}},
  \bibinfo{pages}{171101} (\bibinfo{year}{2013}), \eprint{1306.3983}.

\bibitem[{\citenamefont{Fornengo et~al.}(2014)\citenamefont{Fornengo, Maccione,
  and Vittino}}]{Fornengo:2013xda}
\bibinfo{author}{\bibfnamefont{N.}~\bibnamefont{Fornengo}},
  \bibinfo{author}{\bibfnamefont{L.}~\bibnamefont{Maccione}}, \bibnamefont{and}
  \bibinfo{author}{\bibfnamefont{A.}~\bibnamefont{Vittino}},
  \bibinfo{journal}{JCAP} \textbf{\bibinfo{volume}{1404}}, \bibinfo{pages}{003}
  (\bibinfo{year}{2014}), \eprint{1312.3579}.

\bibitem[{\citenamefont{Maurin et~al.}(2002)\citenamefont{Maurin, Taillet,
  Donato, Salati, Barrau et~al.}}]{Maurin:2002ua}
\bibinfo{author}{\bibfnamefont{D.}~\bibnamefont{Maurin}},
  \bibinfo{author}{\bibfnamefont{R.}~\bibnamefont{Taillet}},
  \bibinfo{author}{\bibfnamefont{F.}~\bibnamefont{Donato}},
  \bibinfo{author}{\bibfnamefont{P.}~\bibnamefont{Salati}},
  \bibinfo{author}{\bibfnamefont{A.}~\bibnamefont{Barrau}},
  \bibnamefont{et~al.} (\bibinfo{year}{2002}), \eprint{astro-ph/0212111}.

\bibitem[{\citenamefont{Di~Mauro
  et~al.}(2014{\natexlab{d}})\citenamefont{Di~Mauro, Donato, Fornengo, Lineros,
  and Vittino}}]{DiMauro:2014iia}
\bibinfo{author}{\bibfnamefont{M.}~\bibnamefont{Di~Mauro}},
  \bibinfo{author}{\bibfnamefont{F.}~\bibnamefont{Donato}},
  \bibinfo{author}{\bibfnamefont{N.}~\bibnamefont{Fornengo}},
  \bibinfo{author}{\bibfnamefont{R.}~\bibnamefont{Lineros}}, \bibnamefont{and}
  \bibinfo{author}{\bibfnamefont{A.}~\bibnamefont{Vittino}},
  \bibinfo{journal}{JCAP} \textbf{\bibinfo{volume}{1404}}, \bibinfo{pages}{006}
  (\bibinfo{year}{2014}{\natexlab{d}}), \eprint{1402.0321}.

\bibitem[{\citenamefont{Ackermann et~al.}(2015)}]{Ackermann:2015tah}
\bibinfo{author}{\bibfnamefont{M.}~\bibnamefont{Ackermann}}
  \bibnamefont{et~al.} (\bibinfo{collaboration}{Fermi-LAT})
  (\bibinfo{year}{2015}), \eprint{1501.05464}.

\bibitem[{\citenamefont{{Press} and {Schechter}}(1974)}]{1974ApJ...187..425P}
\bibinfo{author}{\bibfnamefont{W.~H.} \bibnamefont{{Press}}} \bibnamefont{and}
  \bibinfo{author}{\bibfnamefont{P.}~\bibnamefont{{Schechter}}},
  \bibinfo{journal}{\apj} \textbf{\bibinfo{volume}{187}}, \bibinfo{pages}{425}
  (\bibinfo{year}{1974}).

\bibitem[{\citenamefont{Maccio' et~al.}(2008)\citenamefont{Maccio', Dutton, and
  Bosch}}]{Maccio':2008xb}
\bibinfo{author}{\bibfnamefont{A.~V.} \bibnamefont{Maccio'}},
  \bibinfo{author}{\bibfnamefont{A.~A.} \bibnamefont{Dutton}},
  \bibnamefont{and} \bibinfo{author}{\bibfnamefont{F.~C.~d.}
  \bibnamefont{Bosch}}, \bibinfo{journal}{\mnras}
  \textbf{\bibinfo{volume}{391}}, \bibinfo{pages}{1940} (\bibinfo{year}{2008}),
  \eprint{0805.1926}.

\bibitem[{\citenamefont{{Martinez} et~al.}(2009)\citenamefont{{Martinez},
  {Bullock}, {Kaplinghat}, {Strigari}, and {Trotta}}}]{2009JCAP...06..014M}
\bibinfo{author}{\bibfnamefont{G.~D.} \bibnamefont{{Martinez}}},
  \bibinfo{author}{\bibfnamefont{J.~S.} \bibnamefont{{Bullock}}},
  \bibinfo{author}{\bibfnamefont{M.}~\bibnamefont{{Kaplinghat}}},
  \bibinfo{author}{\bibfnamefont{L.~E.} \bibnamefont{{Strigari}}},
  \bibnamefont{and} \bibinfo{author}{\bibfnamefont{R.}~\bibnamefont{{Trotta}}},
  \bibinfo{journal}{\jcap} \textbf{\bibinfo{volume}{6}}, \bibinfo{eid}{014}
  (\bibinfo{year}{2009}), \eprint{0902.4715}.

\bibitem[{\citenamefont{{Bringmann}}(2009)}]{2009NJPh...11j5027B}
\bibinfo{author}{\bibfnamefont{T.}~\bibnamefont{{Bringmann}}},
  \bibinfo{journal}{New Journal of Physics} \textbf{\bibinfo{volume}{11}},
  \bibinfo{eid}{105027} (\bibinfo{year}{2009}), \eprint{0903.0189}.

\bibitem[{\citenamefont{Sefusatti et~al.}(2014)\citenamefont{Sefusatti,
  Zaharijas, Serpico, Theurel, and Gustafsson}}]{Sefusatti:2014vha}
\bibinfo{author}{\bibfnamefont{E.}~\bibnamefont{Sefusatti}},
  \bibinfo{author}{\bibfnamefont{G.}~\bibnamefont{Zaharijas}},
  \bibinfo{author}{\bibfnamefont{P.~D.} \bibnamefont{Serpico}},
  \bibinfo{author}{\bibfnamefont{D.}~\bibnamefont{Theurel}}, \bibnamefont{and}
  \bibinfo{author}{\bibfnamefont{M.}~\bibnamefont{Gustafsson}},
  \bibinfo{journal}{Mon.Not.Roy.Astron.Soc.} \textbf{\bibinfo{volume}{441}},
  \bibinfo{pages}{1861} (\bibinfo{year}{2014}), \eprint{1401.2117}.

\bibitem[{\citenamefont{{Serpico} et~al.}(2012)\citenamefont{{Serpico},
  {Sefusatti}, {Gustafsson}, and {Zaharijas}}}]{2012MNRAS.421L..87S}
\bibinfo{author}{\bibfnamefont{P.~D.} \bibnamefont{{Serpico}}},
  \bibinfo{author}{\bibfnamefont{E.}~\bibnamefont{{Sefusatti}}},
  \bibinfo{author}{\bibfnamefont{M.}~\bibnamefont{{Gustafsson}}},
  \bibnamefont{and}
  \bibinfo{author}{\bibfnamefont{G.}~\bibnamefont{{Zaharijas}}},
  \bibinfo{journal}{\mnras} \textbf{\bibinfo{volume}{421}},
  \bibinfo{pages}{L87} (\bibinfo{year}{2012}), \eprint{1109.0095}.

\bibitem[{\citenamefont{Ackermann
  et~al.}(2014{\natexlab{c}})}]{Ackermann:2013yva}
\bibinfo{author}{\bibfnamefont{M.}~\bibnamefont{Ackermann}}
  \bibnamefont{et~al.}, \bibinfo{journal}{\prd} \textbf{\bibinfo{volume}{89}},
  \bibinfo{pages}{042001} (\bibinfo{year}{2014}{\natexlab{c}}),
  \eprint{1310.0828}.

\bibitem[{\citenamefont{Aleksić et~al.}(2014)\citenamefont{Aleksić, Ansoldi,
  Antonelli, Antoranz, Babic et~al.}}]{Aleksic:2013xea}
\bibinfo{author}{\bibfnamefont{J.}~\bibnamefont{Aleksić}},
  \bibinfo{author}{\bibfnamefont{S.}~\bibnamefont{Ansoldi}},
  \bibinfo{author}{\bibfnamefont{L.}~\bibnamefont{Antonelli}},
  \bibinfo{author}{\bibfnamefont{P.}~\bibnamefont{Antoranz}},
  \bibinfo{author}{\bibfnamefont{A.}~\bibnamefont{Babic}},
  \bibnamefont{et~al.}, \bibinfo{journal}{JCAP}
  \textbf{\bibinfo{volume}{1402}}, \bibinfo{pages}{008} (\bibinfo{year}{2014}),
  \eprint{1312.1535}.

\bibitem[{\citenamefont{Abramowski et~al.}(2014)}]{Abramowski:2014tra}
\bibinfo{author}{\bibfnamefont{A.}~\bibnamefont{Abramowski}}
  \bibnamefont{et~al.} (\bibinfo{collaboration}{HESS Collaboration})
  (\bibinfo{year}{2014}), \eprint{1410.2589}.

\bibitem[{\citenamefont{{Fornengo} et~al.}(2012)\citenamefont{{Fornengo},
  {Lineros}, {Regis}, and {Taoso}}}]{2012JCAP...01..005F}
\bibinfo{author}{\bibfnamefont{N.}~\bibnamefont{{Fornengo}}},
  \bibinfo{author}{\bibfnamefont{R.~A.} \bibnamefont{{Lineros}}},
  \bibinfo{author}{\bibfnamefont{M.}~\bibnamefont{{Regis}}}, \bibnamefont{and}
  \bibinfo{author}{\bibfnamefont{M.}~\bibnamefont{{Taoso}}},
  \bibinfo{journal}{JCAP} \textbf{\bibinfo{volume}{1}}, \bibinfo{pages}{5}
  (\bibinfo{year}{2012}), \eprint{1110.4337}.

\bibitem[{\citenamefont{{Ackermann} et~al.}(2010)\citenamefont{{Ackermann},
  {Ajello}, {Allafort}, {Baldini} et~al.}}]{2010JCAP...05..025A}
\bibinfo{author}{\bibfnamefont{M.}~\bibnamefont{{Ackermann}}},
  \bibinfo{author}{\bibfnamefont{M.}~\bibnamefont{{Ajello}}},
  \bibinfo{author}{\bibfnamefont{A.}~\bibnamefont{{Allafort}}},
  \bibinfo{author}{\bibfnamefont{L.}~\bibnamefont{{Baldini}}},
  \bibnamefont{et~al.}, \bibinfo{journal}{JCAP} \textbf{\bibinfo{volume}{5}},
  \bibinfo{pages}{25} (\bibinfo{year}{2010}), \eprint{1002.2239}.

\bibitem[{\citenamefont{{Galli} et~al.}(2011)\citenamefont{{Galli}, {Iocco},
  {Bertone}, and {Melchiorri}}}]{2011PhRvD..84b7302G}
\bibinfo{author}{\bibfnamefont{S.}~\bibnamefont{{Galli}}},
  \bibinfo{author}{\bibfnamefont{F.}~\bibnamefont{{Iocco}}},
  \bibinfo{author}{\bibfnamefont{G.}~\bibnamefont{{Bertone}}},
  \bibnamefont{and}
  \bibinfo{author}{\bibfnamefont{A.}~\bibnamefont{{Melchiorri}}},
  \bibinfo{journal}{\prd} \textbf{\bibinfo{volume}{84}},
  \bibinfo{pages}{027302} (\bibinfo{year}{2011}), \eprint{1106.1528}.

\end{thebibliography}

\end{document}